\documentclass[prb,reprint,superscriptaddress,twocolumn]{revtex4-2}
\usepackage{amsmath,amssymb,bm,braket,mathrsfs,mathtools}
\usepackage{multirow}
\usepackage{hyperref}
\usepackage[dvipdfmx]{graphicx}
\usepackage{color}

\DeclareMathOperator{\tr}{Tr}

\begin{document}

\allowdisplaybreaks

\title{Nuclear spin relaxation rate of nonunitary Dirac and Weyl superconductors
}

\author {Koki Maeno}
\affiliation{Department of Applied Physics, Nagoya University, Nagoya, 464-8603, Japan}
\author {Yuki Kawaguchi}
\affiliation{Department of Applied Physics, Nagoya University, Nagoya, 464-8603, Japan}
\author {Yasuhiro Asano}
\affiliation{Department of Applied Physics, Hokkaido University, Supporo, 060-8628, Japan}
\author {Shingo Kobayashi}
\affiliation{RIKEN Center for Emergent Matter Science, Wako, Saitama, 351-0198, Japan}

\begin{abstract}
Nonunitary superconductivity has attracted renewed interest as a novel gapless phase of matter.
In this study, we investigate the superconducting gap structure of nonunitary odd-parity chiral pairing states in a superconductor involving strong spin-orbit interactions. 
By applying a group theoretical classification of chiral states in terms of discrete rotation symmetry, we categorized all possible point-nodal gap structures in nonunitary chiral states into four types in terms of the 
topological number of nodes and node positions relative to the rotation axis.
In addition to conventional Dirac and Weyl point nodes, we identify a novel type of Dirac point node 
unique to nonunitary chiral superconducting states. 
The node type can be identified experimentally based on the temperature dependence of the nuclear magnetic resonance longitudinal relaxation rate.
The implication of our results for a nonunitary odd-parity superconductor in UTe$_2$ is also discussed.

\end{abstract}

\maketitle

\section{introduction}
The striking feature of unconventional superconductors (SCs) is that the internal phase of the pair potential varies depending on the relative direction between the two electrons that form the Cooper pair. Accordingly, sign changes and phase singularities of the pair potential are inevitably accompanied by gap closing points, forming line and point nodes, respectively.
At low temperatures, various physical quantities such as specific heat, magnetic field penetration length, and nuclear magnetic resonance (NMR) relaxation rate exhibit characteristic power-low dependences on temperature $T$, 
reflecting low-energy quasiparticle excitation around the nodes.
For instance, the NMR relaxation rate, which is a powerful experimental probe for detecting low-energy excitations, 
shows $T^5$ ($T^3$) dependence at low temperatures in an SC with linear point nodes (line nodes) in its gap structures~\cite{sigrist1991}. 
In addition, the appearance of the Hebel-Slichter peak below the critical temperature is the result of a fully gapped 
excitation spectrum, suggesting a conventional $s$-wave Cooper pairing ~\cite{hebel1959}.
The internal phase of the pair potential is also a source of the topologically non-trivial 
states in an SC~\cite{qi:rmp2011,tanaka:jpsj2011,sato:jpsj2016,mizushima:jpsj2016,sato:rpp2017,chiu:rmp2016}.
From a topological perspective, the existence of nodes can be well explained by a non-zero topological number. 
Indeed, the bulk-boundary correspondence ensures the existence of zero-energy Andreev bound states at a certain 
surface of unconventional SCs~\cite{sato2011,schnyder2011,brydon2011,yada2011,schnyder2012,matsuura2013,kobayashi14,Kobayashi15PRB,Kobayashi18}. 

In time-reversal symmetry-breaking odd-parity superconducting states,
any line node is fragile owing to inevitable perturbations in real SCs~\cite{Blount1985,kobayashi14}, 
and only point nodes are stable in their gap structures. 
The superfluid $^3$He-A phase is a well-known example of such a state, where 
the gap structure has two Dirac point nodes, point nodes with spin degeneracy. 
$^3$He-A phase is often called a chiral state because Cooper pairs spontaneously have nonzero angular momenta, which 
results in the breakdown of the time-reversal symmetry.
The angular momentum of a Cooper pair is well characterized by the
eigenvalues of the rotation operator around an axis. 
In the presence of continuous rotational symmetry, Dirac point nodes appear on the 
rotation axis in $^3$He-A phase. 

\begin{figure}[tbp]
\centering
\includegraphics[width=8cm]{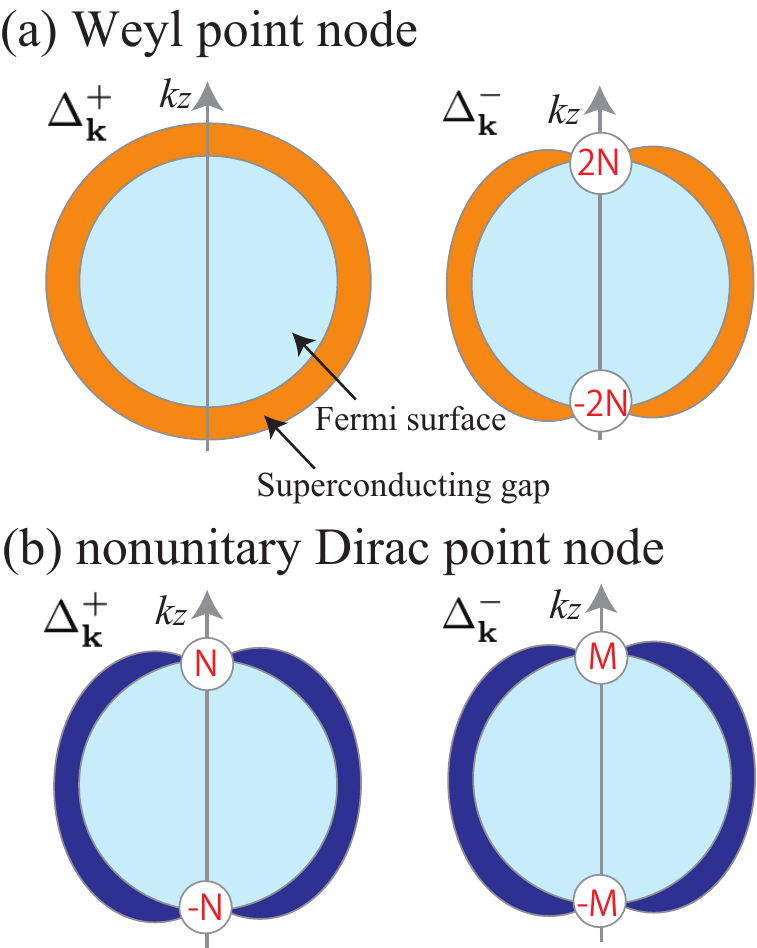}
\caption{ (Color online) Schematic illustration of the gap structures of (a) Weyl point nodes and (b) nonunitary Dirac point nodes ($N,M $ are non-zero integers). The left (right) panel indicates the gap structure of $\Delta_{\bf k}^+$ ($\Delta_{\bf k}^-$) [Eq.~(\ref{eq:gappm})], where nonunitary odd-parity SCs generally satisfy $\Delta_{\bf k}^+ \neq \Delta_{\bf k}^-$. The white circles describe the position of the point node, and the number in the circle indicates the Chern number [Eq.~(\ref{eq:ch})], which is related to the dispersion around the point nodes. The Weyl point nodes are further classified into the ones on [type (i)] and off [type (ii)] axis, whereas the nonunitary Dirac point nodes are categorized to the ones with $|N|=|M|$ [type (iii)] and $|N|\neq |M|$ [type (iv)]. See Fig.~\ref{fig:node1}. 
}
\label{fig:image}
\end{figure}

The concept of chiral superconductivity has recently been applied to superconducting states realized in the presence of strong spin-orbit interactions~\cite{kozii2016,venderbos2018,Sumita19,Ono2022}. 
In materials with strong spin-orbit interactions, an electron is well characterized by a pseudospin that represents a combined degree of freedom 
between the spin and orbital. At the same time, the continuous rotation symmetry breaks into discrete rotation symmetry because of the underlying crystal structures.
By applying group theory, a variety of Weyl point nodes, point nodes without pseudospin degeneracy, have been predicted to appear in odd-parity SCs~\cite{kozii2016}. Schematic pictures of Weyl point nodes are shown in Fig.~\ref{fig:image} (a). 
The low-energy excitations around a Weyl point node depend on the pseudospin of the 
quasiparticle. 
Generally, such superconducting states are chiral superconducting states belonging to the nonunitary class~\cite{sigrist1991}; that is, there is no pseudospin degeneracy in the excitation spectrum owing to the pseudospin-dependent pair potentials.
The physical quantities in nonunitary states would therefore exhibit qualitatively different behaviors from those well-established in unitary superconducting states~\cite{kozii2016}.

In this study, we explore all possible point nodes in nonunitary odd-parity chiral 
superconducting states and analyze their characteristic low-energy excitations.
For this purpose, we
first reclassified the nonunitary odd-parity chiral pairing states in terms of $n$-fold rotation ($C_n$) 
symmetry and Cooper pair angular momentum.
In addition to the conventional Dirac and Weyl point nodes, in this study, we found that a novel type of Dirac point node is possible in a nonunitary chiral SC. 
Figure~\ref{fig:image} (b) shows the gap structures around such a node, which we call a nonunitary Dirac point 
node. 
Although pseudospin degeneracy holds at a nonunitary Dirac point node, the excitations around the node depend on the pseudospin of the electron. 
Based on this classification, we demonstrate that the possible point nodes in the gap structures were categorized into four types in terms of topological numbers and positions in relation to the rotational axis: 
(i) Weyl point nodes along rotation axis, 
(ii) Weyl point nodes off the rotation axis, 
(iii) Nonunitary Dirac point nodes characterized by a Chern number, 
and (iv) nonunitary Dirac point nodes characterized by a pair of Chern numbers. 
We further demonstrate that the difference between the node types appears in the temperature dependence of the NMR longitudinal relaxation rate, $1/T_1$ as summarized in Table~\ref{tab:temp_nmr}. As in the case of unitary pairs with a nodal gap structure, the relaxation rate exhibits power-law behavior: $1/T_1 \sim AT^\gamma$, where $A$ and $\gamma$ are positive real values. However, because the low-energy excitations in nonunitary SCs are strongly dependent on the pseudospin structure, the relaxation rate, i.e., the values of $A$ and $\gamma$, is sensitive to the relative angle $\theta_I$ between the rotation symmetry axis and the external magnetic field. In particular, we found that the point nodes of types (i) and (iv) can be distinguished from others by observing the temperature dependence of $(1/T_1)_\parallel/(1/T_1)_\perp$, where the subscripts $\parallel$ and $\perp$ denote the values at $\theta_I=0$ and $\pi/2$, respectively.
Finally, we discuss the implications of our results for the recently discovered superconductivity in UTe$_2$~\cite{ran2019nearly} 
which is a promising material candidate for a nonunitary odd-parity SC with point 
nodes~\cite{aoki2022,metz2019,kittaka2020,jiao2020chiral,hayes2021,bae2021anomalous,aoki2020,hayes2021,ishihara2021chiral}.

\begin{table*}[tb]
\caption{
Temperature dependence of NMR relaxation rates at low temperature for nonunitary chiral SCs. We categorize the point nodes into four types in terms of the Chern numbers $N,M$ (see Fig.~\ref{fig:image}) and the position relative to the rotation axis: (i) Weyl point nodes on the rotation axis, (ii) Weyl point nodes off the rotation axis, (iii) nonunitary Dirac point nodes with $|N|=|M|$, and (iv) nonunitary Dirac point nodes with $|N| \neq |M|$. We define $(1/T_1)_{\parallel (\perp)}$ as $1/T_1$ with an applied magnetic field being parallel (perpendicular) to the rotation axis. Table~\ref{tab:temp_nmr} shows the temperature dependence of $(1/T_1)_{\parallel}$ and $(1/T_1)_{\perp}$ with $N,M >0$ in comparison with unitary Dirac point node states (the last column).
Each coefficient depends on the details of systems, whose explicit forms are discussed in Secs.~\ref{sec:anisotropicexp} and \ref{sec:anisotropiccoeff}. The last row represents the temperature dependence of a ratio between $(1/T_1)_{\parallel}$ and $(1/T_1)_{\perp}$, where ``const.'' indicates that the ratio is independent of the temperature.
}
\label{tab:temp_nmr}
\begin{tabular}{cccccc}
\hline\hline
Observable & (i) & (ii) & (iii) & (iv) & Unitary \\ \hline 
$\left(\frac{1}{T_1}\right)_{\parallel}$ & $ A_N \,T^{\frac{2}{N}+2}$ & $ D_{\parallel}\, T^{5}$ & $ D_{N,\parallel} \,T^{\frac{4}{N}+1}$ & $ B_{N,M} \,T^{\frac{2}{N}+\frac{2}{M}+1}$ & $ D_N \,T^{\frac{4}{N}+1}$ \\
$\left(\frac{1}{T_1}\right)_{\perp}$ & $ D_{2N} \,T^{\frac{2}{N}+1}$ & $ D_{\perp} \,T^{5}$ & $ D_{N,\perp} \, T^{\frac{4}{N}+1}$ & $ D_{N}^+ \,T^{\frac{4}{N}+1} + D_{M}^- \,T^{\frac{4}{M}+1}$ & $ D_N \,T^{\frac{4}{N}+1}$ \\
$ \frac{(1/T_1)_{\parallel}}{(1/T_1)_{\perp}}$ & $T$ & const. & const. & $T^{\left|\frac{2}{N}-\frac{2}{M}\right|}$ & const. \\
\hline\hline
\end{tabular} 
\end{table*}

The remainder of this paper is organized as follows.
In Sec.~\ref{sec:chiral}, we reclassify the chiral pairing states in terms of $C_n$ symmetry and Cooper pair angular momentum. We categorized all possible point nodes into four types. In Sec.~\ref{sec:spin}, we discuss the power-law dependence of $1/T_1$ on temperature in a unitary chiral SC. In Sec.~\ref{sec:results}, the calculation of $1/T_1$ was applied to nonunitary chiral SCs. 
Finally, we summarize the results and discuss their application to UTe$_2$ in Sec.~\ref{sec:summary}. 
In Appendices~\ref{app:formula}, \ref{app:othergap} and~\ref{app:calculation}, the basic properties of nonunitary odd-parity SCs, some gap structures that are not discussed in the main text, and the calculation of NMR relaxation rates are discussed.
\section{Nonunitary Chiral Cooper pairs}
\label{sec:chiral}
We consider chiral SCs in a system with strong spin-orbit coupling and inversion symmetry ($I$: $I^2=1$). We assume that TR symmetry ($T$: $T^2=-1$) is preserved in the normal state and spontaneously broken in the superconducting states. We focus on odd-parity Cooper pairs, that is, the pair potential satisfies $I\hat{\Delta}_{\bf k}I^{\rm T} =-\hat{\Delta}_{-{\bf k}}$. Owing to Fermi statistics, odd-parity Cooper pairs are pseudospin-triplets on a single-band pseudospin basis. 
Here, we consider electrons with angular momentum $j$, which is a combination of spin and orbital angular momenta, and thus, takes a half integer. We assume that among the Kramers pairs of $|j,j_z\rangle$ and $|j,-j_z\rangle$ $(j_z=1/2, 3/2, \cdots j)$, only the pairs of fixed $j_z$ form Cooper pairs. For simplicity, we consider only the two relevant internal states and use the pseudospin $s=\{\uparrow,\downarrow\}$ description. The pseudospin-triplet pairing states are characterized by the $\bm{d}$ vector as 
\begin{align}
[\hat{\Delta}_{\bf k }]_{ss'} = \bm{d}_{\bf k} \cdot [\bm{\sigma}(i\sigma_y)]_{ss'} , \label{eq:pairpot}
\end{align}
where $\bm{d}_{-\bf k} =-\bm{d}_{\bf k}$ and $\bm{\sigma} = (\sigma_x,\sigma_y,\sigma_z)$ are Pauli matrices in the pseudospin space. Pseudospin-triplet pairings without TR symmetry generally induce nonzero magnetization, $\bm{q}_{\bf k} =\tr(\hat{\Delta}_{\bf k}^{\dagger} \bm{\sigma} \hat{\Delta}_{\bf k} )/2= i \bm{d}_{\bf k} \times \bm{d}^{\ast}_{\bf k}$, which is called the $\bm{q}$ vector~\cite{sigrist1991}.

We are interested in nonunitary chiral pairing states, i.e., $\hat{\Delta}_{\bf k} \hat{\Delta}^{\dagger}_{\bf k}$ is not proportional to the identity matrix~\cite{sigrist1991}. 
The nonunitary properties of pseudospin-triplet pairing states are characterized by the $\bm{q}$ vector. The energy spectrum of superconducting states is described in~\cite{sigrist1991} 
\begin{align}
E_{{\bf k}}^\pm &= \sqrt{\xi_{\bf k }^2 + (\Delta_{\bf k}^{\pm})^2}, \label{eq:enespec} \\ 
\Delta_{\bf k}^{\pm} &\equiv \sqrt{|\bm{d}_{\bf k }|^2 \pm |\bm{q}_{\bf k }|}, \label{eq:gappm}
\end{align}
where $\xi_{\bf k }$ is the normal-state energy relative to the Fermi energy. Thus, $\bm{q}_{\bf k } \neq \bm{0}$ implies that the magnitude of the superconducting gaps depends on the pseudospin structure on the Fermi surface. The nodes appear when $E_{\bf k}^{\alpha} =0$, i.e., $\xi_{\bf k} =0$ and $\Delta_{\bf k}^{\alpha}=0$ ($\alpha = \pm$). The first condition implies that the nodes are on the Fermi surface, whereas the second condition determines the position of the nodes and their pseudospin degeneracy. In the second condition, the Dirac (Weyl) point nodes are defined as points that satisfy $|\bm{d}_{\bf k }|=|\bm{q}_{\bf k }|=0$ ($|\bm{d}_{\bf k }|^2 = |\bm{q}_{\bf k }|\neq 0$).

In the following, we reclassify the nonunitary chiral pairing states in terms of $C_n$ symmetry. The operation of the $n$-fold rotation about the $z$ axis changes the annihilation operator of an electron with momentum ${\bf k}$ and projected angular momentum $j_z$ as follows:
\begin{align}
c_{{\bf k}, j_z} \to c_{R_n{\bf k},j_z} e^{-i \frac{2\pi}{n} j_z}, \label{eq:rot_c}
\end{align}
where $R_n$ is the $n$-fold rotation in momentum space, for example, $R_4: (k_x,k_y,k_z) \to (-k_y,k_x,k_z)$, and $j_z$ determines the phase change associated with the rotation in pseudospin space. On the rotation symmetric line satisfying $R_n{\bf k}={\bf k}$, the electronic states indicate the eigenstates of the $C_n$ operations and their eigenvalues depend on $j_z$. Thus, under $C_n$ symmetry, the electronic states are classified as $j_z$.

When Cooper pairs have the projected angular momentum $J_z$ $(0\le J_z\le 2j_z)$, the gap function after the $n$-fold rotation should satisfy~\cite{sato:jpsj2016}
\begin{align}
\hat{C}_{n,j_z} \hat{\Delta}_{\bf k} \hat{C}_{n,j_z}^{T}= e^{-i \frac{2\pi}{n} J_z} \hat{\Delta}_{R_n{\bf k}}, \label{eq:gap_Cn} 
\end{align}
with $\hat{C}_{n,j_z} = {\rm diag} [e^{-i \frac{2\pi}{n} j_z} , e^{i \frac{2\pi}{n} j_z} ] $. 
By using the $\bm{d}$vectors, Eq.~(\ref{eq:gap_Cn}) can be rewritten as
\begin{align}
e^{-i\frac{4\pi}{n} j_z} d_{{\bf k}}^- &= e^{-i \frac{2\pi}{n} J_z} d_{ R_n{\bf k}}^-, \label{eq:dp_const}\\
e^{i\frac{4\pi}{n} j_z} d_{ {\bf k}}^+ &= e^{-i \frac{2\pi}{n} J_z} d_{ R_n{\bf k}}^+, \label{eq:dm_const}\\
d_{ {\bf k}}^z &= e^{-i \frac{2\pi}{n} J_z} d_{ R_n{\bf k}}^z, \label{eq:dz_const}
\end{align}
where $d_{{\bf k}}^\pm = d_{ {\bf k}}^x \pm i d_{ {\bf k}}^y$, In addition, we expand $\bm{d}_{\bf k} $ to the leading order of ${\bf k}$, $ d_{ {\bf k}}^i \propto k_+^{p_i} k_-^{q_i} k_z^{r_i} $ ($i=+,-,z$), where $p_i,q_i,r_i$ are nonnegative integers, and $k_{\pm} \equiv k_x \pm i k_y$ satisfies $R_nk_{\pm} = e^{\pm i\frac{2\pi}{n}}k_{\pm}$. Substituting these into Eqs.~(\ref{eq:dp_const}), (\ref{eq:dm_const}), and (\ref{eq:dz_const}), we obtain the relations between $J_z$, $j_z$, $p_i$, and $q_i$:
\begin{align}
2j_z + p_- -q_- &= J_z \; \mod n, \label{eq:relpqj1}\\
-2j_z + p_+ -q_+ &= J_z \; \mod n, \label{eq:relpqj2}\\
p_z-q_z &= J_z \; \mod n, \label{eq:relpqj3}
\end{align}
which determines the momentum dependence of the $\bm{d}$ vector perpendicular to the rotational axis. In addition, $r_i$ is constrained by inversion symmetry; because the $\bm{d}$ vector is an odd function of ${\bf k}$, we require that $p_i+q_i+r_i$ is an odd integer. The symmetry-allowed forms of $\bm{d}$ vectors are summarized in Table~\ref{tab:form_factors}, where we only consider the lowest-order terms of ${\bf k}$ in each $d^i_{\bf k}$ ($i=+,-,z$). Because Cooper pairs with $J_z<0$ are related to those with $J_z>0$ by TR symmetry, we assume $J_z > 0$ in Table~\ref{tab:form_factors}.

We determined the gap structure of these pair potentials by analyzing Eq.~(\ref{eq:enespec}). Dirac point nodes require $|\bm{d}_{\bf k }|=0$, that is, $d_{{\bf k}}^+=d_{{\bf k}}^-=d_{{\bf k}}^z=0$, which automatically leads to $|\bm{q}_{\bf k}|=0$. It follows that the Dirac point nodes are on the rotational axis. It should be noted that even though both $E_{\bf k}^+$ and $E_{\bf k}^-$ vanish at Dirac point nodes, their dispersions are not necessarily degenerate. However, the Weyl point nodes can be on or off the rotation axis.

To characterize the point nodes, we employed a topological argument. The Chern number $Q_{\alpha}$ on sphere $S^2$ enclosing a point node is defined by
\begin{align}
Q_{\alpha} = \frac{1}{2\pi} \int_{S^2} d{\bf k} \; \epsilon_{ij} \partial_{k_i} A_{{j} {\bf k}}^{\alpha}, \label{eq:ch}
\end{align}
with the antisymmetric tensor $\epsilon_{ij} = -\epsilon_{ji} $ ($|\epsilon_{ij}|=1$) and Berry connection $A_{{i} {\bf k}}^{\alpha} \equiv i\langle u_{{\bf k} \alpha}| \partial_{k_i} u_{{\bf k}\alpha} \rangle$. Here, $|u_{{\bf k} \alpha} \rangle$ is the eigenvector with eigenvalue $-E_{{\bf k}}^{\alpha} $ of the Bogoliubov-de Gennes (BdG) Hamiltonian, the properties of which are summarized in Appendix~\ref{app:formula}. The concrete form of $|u_{{\bf k}\alpha}\rangle$ is given by (\ref{eq:ustate}) and (\ref{eq:vstate}). We assume that the energy spectra are non-degenerate on $S^2$, which is satisfied for nonunitary point node states. $|Q_{\alpha}|$ is related to the dispersion relation around point nodes in $E_{\bf k}^{\alpha}$. The details are presented in Sec.~\ref{sec:node_typei}-\ref{sec:node_typevi}.
The Chern number is conserved on the sphere even under a small perturbation that causes a point node in the rotation axis to split into multiple off-axis nodes as long as the sphere encloses the split point nodes. This conservation law helps us understand the structure of the split point nodes, where the sum of all Chern numbers in the Brillouin zone is zero. In addition, a nonzero Chern number manifests itself in the existence of surface zero energy states via bulk-boundary correspondence, which provides further evidence of nonunitary chiral SCs~\cite{kozii2016}. 

In Table~\ref{tab:form_factors}, we categorize the gap structures into four types in terms of the Chern number and the location relative to the rotation axis: (i) Weyl point nodes on the rotation axis, (ii) Weyl point nodes off the rotation axis, (iii) nonunitary Dirac point nodes with $|Q_+|=|Q_-|$, and (iv) nonunitary Dirac point nodes with $|Q_+| \neq |Q_-|$.

Note that the classification in Table~\ref{tab:form_factors} corresponds to that in Table~2 in Ref.~\onlinecite{kozii2016}, where the gap structures around the Weyl point nodes [(i) and (ii)] in Table~\ref{tab:form_factors} are discussed. In our study, we also focus on the nonunitary Dirac point nodes [(iii) and (iv)] and examine their properties. 

\section{Nonunitary point node states}

In the following, we present concrete examples of gap structures of types (i)--(iv). Schematics of the gap structures are shown in Fig.~\ref{fig:node1}. Focusing on the dispersion relation around the point nodes in $\Delta^{\pm}_{\bf k}$, we examine the relation between the dispersion relation and Chern number $Q_{\alpha}$. In the low-energy regime, the energy spectrum is approximated around the point node, which is described by the momentum relative to the point nodes, ${\bf p} \equiv {\bf k} - {\bf k}_0$ $(|{\bf p}| \ll k_F$), where $k_F$ is the Fermi wavenumber and ${\bf k}_0$ is the momentum at the point node, i.e., ${\bf k}_0 = {\bf k}_{\pm} \equiv (0,0,\pm k_F)$ when the node is on the rotation axis. To simplify the notation, we define $\hat{k}_i \equiv k_i/k_F$ ($i=x,y,z$), $\hat{k}_{\pm} \equiv \hat{k}_x\pm i \hat{k}_y$ and $\hat{k}_{\perp} = \sqrt{\hat{k}_x^2+\hat{k}_y^2}$. The same notation is applied to ${\bf p}$.

\begin{table}[tb]
\caption{
Classification of pseudospin-triplet pairing states with $J_z \neq 0$ under the $C_n$ symmetry. Table~\ref{tab:form_factors} shows the ${\bf k}$ dependence of the $\bm{d}$-vector in the leading order for various sets of $n$, $J_z$, and $j_z$, where $J_z$ and $\pm j_z$ are the projected angular momentum of a Cooper pair and those of electrons constituting the Cooper pair, $k_{\pm} = k_x\pm ik_y$, and $d_{ {\bf k}}^\pm = d_{ {\bf k}}^x \pm i d_{ {\bf k}}^y$. Here, we only consider the cases of $ j_z=1/2, 3/2, 5/2$ and $J_z=1, 2, 3$. The results for $J_z<0$ are obtained from those for $|J_z|$ via the TR symmetry. The seventh and eighth columns indicate the node types, Weyl (W) or Dirac (D), and types (i)--(iv). The double labels, e.g., (i, ii), imply the coexistence of on and off axis point nodes. The last two columns indicate the Chern number, $|Q_\pm|$, of the point nodes on the rotation axis in $E^{\pm}_{\bf k}$. Here, ``$-$'' refers to the absence of point nodes on the rotation axis.}
\label{tab:form_factors}
\begin{tabular}{cccccccccc}
\hline\hline
$n$ & $J_z$ & $j_z$ & $d_{{\bf k}}^-$ & $d_{{\bf k}}^+$ & $d_{{\bf k}}^z$& Node & Types &$|Q_+|$&$|Q_-|$\\
\hline 
$2$ & $1$ & $\frac{1}{2}$ &$k_z$ & $k_z$ & $k_+$, $k_-$& W & (ii) & $-$ & $-$\\
$3$ & $1$ & $\frac{1}{2}$ &$k_z$ & $k_-$ & $k_+$ & W, W & (i, ii)& $-$ & $1$ \\
$3$ & $1$ & $\frac{3}{2}$ &$k_+$ & $k_+$ & $k_+$ & D & (iii)& $1$ & $1$\\
$3$ & $1$ & $\frac{5}{2}$ &$k_-$ & $k_z$ & $k_+$ & W, W & (i, ii)& $-$ & $1$\\
$4$ & $1$ & $\frac{1}{2}$ &$k_z$ & $k_zk_{\pm}^2$ & $k_+ $ & W & (i) & $-$ & $ 2$\\
$4$ & $1$ & $\frac{3}{2}$ &$k_zk_{\pm}^2$ & $k_z$ & $k_+$ & W & (i) & $-$ & $ 2$ \\
$4$ & $2$ & $\frac{1}{2}$ &$k_+$ & $k_-$ & $k_zk_{\pm}^2$ &D & (iii) & $1$ & $1$\\
$4$ & $2$ & $\frac{3}{2}$ &$k_-$ & $k_+$ & $k_zk_{\pm}^2$ & D & (iii) & $1$ & $1$ \\
$6$ & $1$ & $\frac{1}{2}$ &$k_z$ & $k_zk_+^2$ & $k_+$ & W & (i) & $-$ & $2$ \\
$6$ & $1$ & $\frac{3}{2}$ &$k_zk_-^2$ & $k_zk_-^2$ & $k_+$ & D & (iii) & $1$ & $1$\\
$6$ & $1$ & $\frac{5}{2}$ &$k_zk_+^2$ & $k_z$ & $k_+$ & W & (i) & $-$ & $2$ \\
$6$ & $2$ & $\frac{1}{2}$ &$k_+$ & $k_{\pm}^3$ & $k_zk_+^2$ & D & (iv) & $1$ & $ 3$\\
$6$ & $2$ & $\frac{3}{2}$ &$k_-$ & $k_-$ & $k_zk_+^2$ & D & (iii) & $1$ & $1$ \\
$6$ & $2$ & $\frac{5}{2}$ &$k_{\pm}^3$ & $k_+$ & $k_zk_+^2$ & D & (iv) & $ 1$ & $3$\\
$6$ & $3$ & $\frac{1}{2}$ &$k_zk_+^2$ & $k_zk_-^2$ & $k_{\pm}^3$ & W, D & (ii, iii) & $2$ & $2$\\
$6$ & $3$ & $\frac{3}{2}$ &$k_z$ & $k_z$ & $k_{\pm}^3$ & W & (ii) & $-$ & $-$ \\
$6$ & $3$ & $\frac{5}{2}$ &$k_zk_-^2$ & $k_zk_+^2$ & $k_{\pm}^3$ & W, D & (ii, iii) & $2$ & $2$ \\
\hline\hline
\end{tabular} 
\end{table} 

\subsection{Type (i)}
\label{sec:node_typei}
\begin{figure}[tbp]
\centering
\includegraphics[width=8 cm]{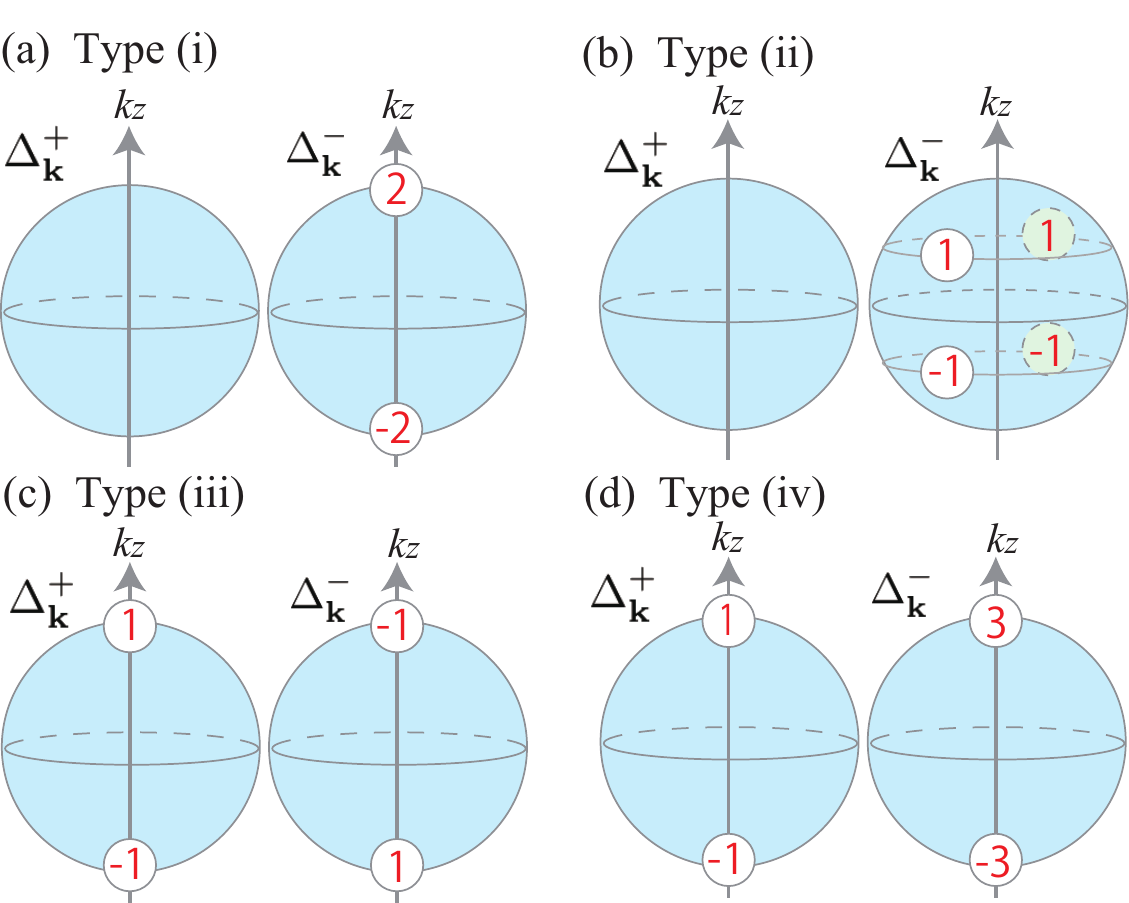}
\caption{ (Color online) Gap structures $\Delta_{\bf k}^{\pm}$ of nonunitary Dirac and Weyl SCs under $C_n$ symmetry. (a), (b), (c), and (d) concretely exhibit the gap structures of types (i)-(iv), where we adopt the $\bm{d}$ vectors of $(n,J_z,j_z) = (6,1,1/2)$, $(2,1,1/2)$, $(4,2,1/2)$, and $(6,2,1/2)$, respectively. Here, the blue sphere represents the Fermi sphere and the white (light-green) circles the position of point nodes at this side (the other side) of the paper. The number in the circle indicates the Chern number.
}
\label{fig:node1}
\end{figure}

We begin with the Weyl point nodes on the rotation axis. For instance, we consider the $\bm{d}$ vector of $(n,J_z,j_z)=(6,1,1/2)$, whose components, $\bm{d}_{\bf k}=(d_{{\bf k}}^x,d_{{\bf k}}^y,d_{{\bf k}}^z)$, are given by
\begin{align}
\bm{d}^{[6,1,\frac{1}{2}]}_{\bf k} = (\lambda_a \hat{k}_z, i \lambda_a \hat{k}_z, \lambda_b \hat{k}_+), \label{eq:dc6}
\end{align}
where $\lambda_a, \lambda_b \in \mathbb{R}$. We neglect the $k$-cubic terms $d_{\bf k}^-\propto k_z k_+^2$ because their contribution to the nodes is small. The $\bm{d}$ vector in Eq.~(\ref{eq:dc6}) leads to the $\bm{q}$ vector as
\begin{align}
\bm{q}^{[6,1,\frac{1}{2}]}_{\bf k} = -2 (\lambda_a \lambda_b \hat{k}_x \hat{k}_z,\lambda_a \lambda_b \hat{k}_y \hat{k}_z, - \lambda_a^2 \hat{k}_z^2), \label{eq:qc6}
\end{align}
which represents the pseudospin structure of the nonunitary pairing state. 
Substituting Eqs.~(\ref{eq:dc6}) and (\ref{eq:qc6}) into Eq.~(\ref{eq:gappm}), the dispersion relation up to the leading order of $\hat{{\bf p}}$ is obtained as
\begin{align}
\Delta_{{\bf k}_\pm+{\bf p} }^+ \simeq2 |\lambda_a|, \ \ \Delta_{{\bf k}_\pm+{\bf p} }^- \simeq \tilde{v}_{\Delta} \hat{p}_{\perp}^2, \label{eq:dis-typei}
\end{align}
where $\tilde{v}_{\Delta} = \lambda_b^2/2|\lambda_a|$, Thus, the point node appears only in $E_{\bf k}^-$ and exhibits quadratic dispersion. See Fig.~\ref{fig:node1} (a) and Eq.~(\ref{eq:ch}) around the point node at the north (south) pole is $Q_-=2 (-2)$. 

In general, the $\bm{d}$ vector of type (i) can be expressed as
\begin{align}
\bm{d}_{\bf k}^{\pm N} &= (\lambda_a \hat{k}_z, - i \lambda_a \hat{k}_z, \lambda_b \hat{k}_{\pm}^N ), \label{eq:generalweyl}
\end{align}
where $N$ is a positive integer~\footnote{When $N$ is an even integer, $\lambda_b \hat{k}_+^N$ is replaced with $\lambda_b \hat{k}_z\hat{k}_+^N$ because of the inversion symmetry}. The corresponding dispersion relation and Chern number are given by $\Delta_{{\bf k}_\pm+{\bf p} }^- \simeq \tilde{v}_{\Delta} \hat{p}_{\perp}^{2N}$ and $Q_- = \pm 2N (\mp 2N)$ at the north (south) pole.

\subsection{Type (ii)}
\label{sec:node_typeii}
Weyl point nodes on the rotation axis are often split into multiple off-axis point nodes via symmetry reduction. To observe the splitting of nodes, we consider the case of $(n,J_z,j_z)=(2,1,1/2)$, which is obtained by reducing the symmetry of $(n,J_z,j_z)=(6,1,1/2)$ from $C_6$ to $C_2$. The corresponding $\bm{d}$ vector contains additional terms in Eq.~(\ref{eq:dc6}), and is given by
\begin{align}
\bm{d}^{[2,1,\frac{1}{2}]}_{\bf k} =&[(\lambda_a +\lambda_c) \hat{k}_z, i (\lambda_a -\lambda_c) \hat{k}_z, \lambda_b k_++\lambda_d \hat{k}_-] \nonumber \\
=& \bm{d}^{[6,1,\frac{1}{2}]}_{\bf k}+(\lambda_c \hat{k}_z, -i \lambda_c \hat{k}_z, \lambda_d \hat{k}_-), \label{eq:dc2}
\end{align}
where $\lambda_a, \lambda_b, \lambda_c, \lambda_d \in \mathbb{R}$. Accordingly, the $\bm{q}$ vector changes from Eq.~(\ref{eq:qc6}) to 
\begin{align}
\bm{q}^{[2,1,\frac{1}{2}]}_{\bf k} =-2\Big(&(\lambda_a-\lambda_c)(\lambda_b+\lambda_d)\hat{k}_x\hat{k}_z, \nonumber \\
&(\lambda_a+\lambda_c)(\lambda_b-\lambda_d)\hat{k}_y\hat{k}_z, \nonumber \\
&-(\lambda_a^2-\lambda_c^2)\hat{k}_z^2\Big). \label{eq:qc2}
\end{align}

To check the dispersion relation, we set $\lambda_d=0$ and $\lambda_a \lambda_c >0$. In this case, by solving $\Delta_{\bf k}^- =0$ ($\Delta_{\bf k}^+$ is fully gapped), the position of the point nodes is represented as
\begin{align}
{\bf k}_{\eta_y,\eta_z} = k_F \left(0, \frac{2 \eta_y \sqrt{\lambda_a \lambda_c}}{\sqrt{\lambda_b^2+4 \lambda_a \lambda_c}}, \frac{\eta_z \lambda_b}{\sqrt{\lambda_b^2+4 \lambda_a \lambda_c}}\right), \label{eq:offnodeC2}
\end{align}
where $\eta_y, \eta_z = \pm$ indicate the position of the point nodes.
Around the point nodes, the energy spectrum is approximated to the leading order of $\hat{{\bf p}}$ as follows:
\begin{align}
\Delta_{{\bf k}_{\eta_y,\eta_z}+{\bf p} }^- \simeq \sqrt{v_{\Delta x}^2 \hat{p}_x^2 + v_{\Delta y}^2 \hat{p}_y^2} , \label{eq:enec2}
\end{align}
where $v_{\Delta x} \equiv 2 \lambda_b \sqrt{\lambda_a \lambda_c}/(\lambda_a +\lambda_c)$ and $v_{\Delta y} \equiv 2 \sqrt{\lambda_a \lambda_c(\lambda_b^2+4 \lambda_a \lambda_c)}/(\lambda_a +\lambda_c)$. Here, we rotate the coordinates to eliminate $\hat{p}_z$ dependence (see Appendix~\ref{app:off}). Thus, the point node has linear dispersion, and the Chern number of each split point node is $|Q_-|=1$. The sign of the Chern numbers obeys the conservation law of the Chern number. 

In general, a $C_n$ symmetric system has at least $2n$ off-axis Weyl point nodes in $C_n$ symmetric systems. The positions of the off-axis Weyl point nodes in $C_3$ and $C_6$ symmetric systems are discussed in Appendix~\ref{app:othergap}.

\subsection{Type (iii)}
\label{sec:node_typeiii}
Nonunitary Dirac point nodes of type (iii) share common properties with chiral (helical) pairing states in unitary SCs that host Dirac point nodes with non-zero (zero) Chern numbers. The nonzero $\bm{q}$ vector of type (iii) originates from the mismatch of coefficients between $d_{{\bf k}}^+$ and $d_{{\bf k}}^-$. To observe this, we consider the gap structure of $(n,J_z,j_z)=(4,2,1/2)$, whose $\bm{d}$ vector and $\bm{q}$ vector are given by
\begin{align}
\bm{d}^{[4,2,\frac{1}{2}]}_{\bf k} &= (\lambda_a \hat{k}_+ +\lambda_b \hat{k}_-, i \lambda_a \hat{k}_+-i \lambda_b \hat{k}_-, 0), \label{eq:dc4} \\
\bm{q}^{[4,2,\frac{1}{2}]}_{\bf k} &= (0,0,2(\lambda_a^2-\lambda_b^2)\hat{k}_{\perp}^2), \label{eq:qc4}
\end{align}
where we neglect the contribution of the $k$-cubic term. The dispersion relation is given by the linear dispersion as
\begin{align}
\Delta_{{\bf k}_\pm+{\bf p} }^+ \simeq v_{\Delta}^+ \hat{p}_{\perp}, \ \ \Delta_{{\bf k}_\pm+{\bf p}}^- \simeq v_{\Delta}^- \hat{p}_{\perp}, \label{eq:gap4212}
\end{align}
with $v_{\Delta}^+ = 2 |\lambda_a|$ and $v_{\Delta}^- = 2 |\lambda_b|$. The corresponding Chern number of $E_{\bf k}^{\pm}$ at the north (south) pole is $Q_{\pm}=\pm1(\mp1)$. The gap structure is shown in Fig.~\ref{fig:node1} (c).

Clearly, the $\bm{d}$ vector is reduced to a helical pairing state when $\lambda_a=\lambda_b$. Note that a helical pairing state preserves TR symmetry; however, the nonunitary Dirac point node state breaks the TR symmetry because $\lambda_a \neq \lambda_b$. Similarly, a nonunitary version of chiral pairing states appears in $(n,J_z,j_z)=(6,2,3/2)$, whose $\bm{d}$-vector is given by replacing $\lambda_a k_+$ with $\lambda_a k_-$ in Eq.~(\ref{eq:dc4}).

For later convenience, we generalize Eq.~(\ref{eq:dc4}) for the cases with higher Chern numbers. The generalized form of the $\bm{d}$ vector can be described by 
\begin{align}
\bm{d}^{N ,\pm M}_{{\bf k}} &= (\lambda_a \hat{k}_{+}^N+\lambda_b \hat{k}_{\pm}^{M}, i \lambda_a \hat{k}_{+}^N-i \lambda_b \hat{k}_{\pm}^{M},0), \label{eq:generalchiral}
\end{align} 
where $N,M$ is a positive integer~\footnote{When $N$ ($M$) is an even integer, the inversion symmetry requests that $\lambda_a \hat{k}_{+}^N$ ($\lambda_b \hat{k}_{\pm}^{M}$) changes to $\lambda_a \hat{k}_z \hat{k}_{+}^N$ ($\lambda_b \hat{k}_z\hat{k}_{\pm}^{M}$)}.
Here, the nonunitary Dirac point nodes of type (iii) satisfy $N=M$.
The dispersion relation and the Chern numbers are given by
\begin{align}
\Delta_{{\bf k}_\pm+{\bf p}}^+ \simeq v_{\Delta}^+ \hat{p}_{\perp}^{N}, \ \ \Delta_{{\bf k}_\pm+{\bf p}}^- \simeq v_{\Delta}^- \hat{p}_{\perp}^{M}, \label{eq:gaptypeiv}
\end{align}
and $Q_+=N$ and $Q_-=\pm M$ at the north pole. Note that a gap structure with $N \neq M$ is also possible, which is categorized as type (iv); the details are discussed in the next section. 

\subsection{Type (iv)}
\label{sec:node_typevi}
Finally, we consider nonunitary Dirac point nodes of type (iv). 
As an example, we focus on the case of $(n,J_z,j_z)=(6,2,1/2)$, where the $\bm{d}$ vector is given by
\begin{align}
\bm{d}_{\bf k}^{[6,2,\frac{1}{2}]} = &(\lambda_a \hat{k}_+ + \lambda_b \hat{k}_+^3+\lambda_c \hat{k}_-^3, \nonumber \\
& i(\lambda_a \hat{k}_+ - \lambda_b \hat{k}_+^3-\lambda_c \hat{k}_-^3),\lambda_d k_z \hat{k}_+^2). \label{eq:dvec611}
\end{align}
Here, we consider the third-order terms of ${\bf k}$ because $E_{{\bf k}}^-$ is completely gapless when $\lambda_b=\lambda_c=\lambda_d=0$. Adding $k$-cubic terms changes the surface node to a point node in $E_{{\bf k}-}$. 

To verify the dispersion relation, we set $\lambda_c=0$ for brevity. Keeping the leading contribution of $\hat{{\bf p}}$, the gap structure around the point nodes is given by
\begin{align}
\Delta_{{\bf k}_\pm+{\bf p} }^+ \simeq v_{\Delta}^+ \hat{p}_{\perp}, \ \ \Delta_{{\bf k}_\pm+{\bf p}}^- \simeq \tilde{v}_{\Delta}^{-} \hat{p}_{\perp}^3 , \label{eq:gap6212}
\end{align}
where $\tilde{v}_{\Delta}^{-}= |2 \lambda_b + \lambda_d^2/2\lambda_a|$. The Chern numbers of the point nodes at the north (south) pole are $Q_+=1(-1)$ and $Q_-=3(-3)$. The gap structure is shown in Fig.~\ref{fig:node1} (d). 
The general form of type (iv) can be described by Eq.~(\ref{eq:generalchiral}), where the dispersion relations satisfy Eq.~(\ref{eq:gaptypeiv}) with $N \neq M$.

\section{NMR relaxation rate}
\label{sec:spin}
We consider the spin-lattice relaxation rate $1/T_1$ of NMR to be a probe for the pseudospin-dependent low-energy density of states. The measurement of $1/T_1$ at low temperatures is a powerful experimental technique for detecting the node structures. The low-temperature power-law behavior of $1/T_1$ was used as a measure of the density of states and enabled the identification of gap structures from the temperature exponent.

We assume that the NMR relaxation originates from the interaction between the nuclear spin and quasiparticle states of the SC, mediated by the hyperfine coupling Hamiltonian between the nuclear and itinerant electrons,
\begin{align}
H_{\rm hf} = \gamma_{\rm n} A_{\rm hf} \sum_{s, s'} \sum_{{\bf k},{\bf k}'} \bm{I} \cdot c_{{\bf k},s}^{\dagger}\bm{\sigma}_{ss'} c_{{\bf k}' , s'}, \label{eq:Hhf}
\end{align}
where $\bm{I}$ denotes the nuclear spin operator vector, $\gamma_{\rm n}$ denotes the nuclear gyromagnetic ratio, and $A_{\rm hf}$ denotes the hyperfine coupling constant. Here, we simplify the hyperfine coupling constant by neglecting the material-dependent momentum dependence and anisotropy so that we extract a universal feature of low-energy excitations around the point nodes.

Using Eq.~(\ref{eq:Hhf}), we examined the NMR longitudinal relaxation rate for nonunitary chiral SCs. For simplicity, we consider the case of nuclear spin of $I=1/2$. Then, $1/T_1$ is formulated using Fermi's golden rule~\cite{hebel1959} as
\begin{align}
\frac{1}{T_1} = \frac{2 \pi}{\hbar} \sum_{\alpha, \beta} \sum_{{\bf k},{\bf k}'} & |\langle -\bm{I}; {\bf k} \alpha | H_{\rm hf} | \bm{I}; {\bf k}' \beta \rangle|^2 f_{{\bf k}'}^\beta (1-f_{\bf k}^\alpha) \nonumber \\
&\times \delta(E_{{\bf k}}^\alpha - E_{{\bf k}'}^\beta-\hbar \omega_0), \label{eq:NMR}
\end{align}
where $\omega_0$ is the NMR frequency, and $ | \bm{I}; {\bf k} \alpha \rangle \equiv | \bm{I} \rangle \otimes | {\bf k} \alpha \rangle$. Here, $ | \bm{I} \rangle$ is the eigenstate of the nuclear spin $\bm{I}$ parallel to the external magnetic field,
and $ | {\bf k} \alpha \rangle$ is the eigenstate of the BdG Hamiltonian with the eigenvalue $E_{{\bf k}}^\alpha$. The explicit form of $ | {\bf k} \alpha \rangle$ is provided in Appendix~\ref{app:formula}. When an external magnetic field is applied in the direction specified by the polar and azimuthal angles $\theta_I$ and $\phi_I$ relative to the symmetry axis, $| \bm{I} \rangle $ is given by 
\begin{align}
| \bm{I} \rangle =(\cos(\theta_{I}/2),e^{i \phi_{I}} \sin(\theta_{I}/2))^T. \label{eq:eigen_nuclear}
\end{align} 
We define $(1/T_1)_{\parallel (\perp)}$ as $1/T_1$ with a fixed angle $\theta_I=0$ ($\theta_I=\pi/2$). The temperature dependence of $1/T_1$ is included in the Fermi-Dirac distribution function $f_{\bf k}^\alpha = (e^{E_{\bf k }^\alpha/k_{\rm B}T}+1)^{-1}$~\footnote{We assume that the temperature dependence of the pair potential can be neglected at low temperatures.}. We study excitations around the point nodes in a low-energy regime characterized by $\hbar \omega_0 \ll k_{\rm B} T \ll \Delta_0$, where $\Delta_0$ is the energy scale of the superconducting gap characterizing the point nodes, for example, $\Delta_0 = {\rm min}[\lambda_a,\lambda_b,\lambda_c,\lambda_d]$ for $\bm{d}^{[2,1,\frac{1}{2}]}_{\bf k}$. Throughout this study, we approximate $\hbar \omega_0 \approx 0$ and use the unit $\hbar =k_{\rm B}=1$.

Before discussing our results, we revisit the power-law temperature dependence of unitary chiral SCs whose energy spectrum is described as
\begin{align}
E_{\bf k} &= \sqrt{\xi_{\bf k}^2 + |\bm{d}_{\bf k}|^2}, \label{eq:eigenuni}\\
\bm{d}_{\bf k} &= \lambda (0,0,\hat{k}_+^N),
\end{align}
where $\lambda \in \mathbb{R}$, $N$ denotes a positive integer, and $\bm{q}_{\bf k}=0$. 
In this case, Eq.~(\ref{eq:NMR}) is reduced to
\begin{align}
\frac{1}{T_1} = \pi \gamma_{\rm n}^2 A_{\rm hf}^2 \sum_{ {\bf k}, {\bf k}'} f_{{\bf k}'} (1-f_{\bf k}) \delta(E_{{\bf k}} - E_{{\bf k}'}), \label{eq:NMRunitary}
\end{align}
where we perform the sum over pseudospins (Appendix \ref{app:nmruni} for details). Thus, $1/T_1$ in unitary chiral SCs is independent of the magnetic field direction. To proceed with the calculation, we assume that $\xi_{\bf k}$ has a spherical Fermi surface, that is, $\xi_{\bf k} =({\bf k}^2-k_F^2)/2m$, where $m$ is the effective mass of the electron. Because the contribution from the low-energy excitations around the point nodes is dominant at low temperatures, the energy spectrum can be approximated around the point nodes at ${\bf k}_{\pm}$ as 
\begin{align}
E_{{\bf k}_\pm + {\bf p}} \simeq \sqrt{v_F^2 \hat{p}_z^2+ v_{\Delta}^2 \hat{p}_{\perp}^{2N} }, \label{eq:EqgenU}
\end{align} 
where $v_{F}=v_F/m$ and $v_{\Delta}=\lambda$.
We replace the summation with the integral $\sum_{\bf k} \to \int_0^{\infty} D(E) dE$, where $D(E)$ is the density of states of the quasiparticles 
\begin{align}
D(E) &= \frac{1}{(2\pi)^3} \int_{\mathbb{R}^3} d^3 p \; \delta (E-E_{{\bf k}_\pm+{\bf p}} ). \label{eq:dos}
\end{align}
Plugging Eq.~(\ref{eq:EqgenU}) into Eq.~(\ref{eq:dos}), we obtain
\begin{align}
D(E) = d_N E^{\frac{2}{N}} \label{eq:dosch}
\end{align}
with
\begin{align}
d_N \equiv \frac{\sqrt{\pi}}{(2\pi)^2} \frac{k_F^3}{N v_F v_{\Delta}^{\frac{2}{N}}} \frac{\Gamma\left(\frac{1}{N}\right)}{\Gamma\left(\frac{N+2}{2N}\right)}. \label{eq:doscoeff}
\end{align}
Consequently, Eq.~(\ref{eq:NMRunitary}) becomes
\begin{align}
\frac{1}{T_1} = \pi \gamma_{\rm n}^2 A_{\rm hf}^2 \int_0^{\infty} D^2(E) \left(- T \frac{\partial f_E }{\partial E} \right) dE, \label{eq:NMRunitaryE}
\end{align}
where $f_{E} (1-f_{E}) = - T \partial f_E/\partial E$, Hence, the temperature dependence originates from the $E$ dependence of the density of states, whose exponent is related to the Chern number of the point nodes through the dispersion relation. 
Using Eqs.~(\ref{eq:dosch}) and (\ref{eq:NMRunitaryE}), $1/T_1$ is calculated as
\begin{align}
\frac{1}{T_1} = \pi \gamma_{\rm n}^2 A_{\rm hf}^2 D_N T^{\frac{4}{N}+1}, \label{eq:Tunitary}
\end{align}
with 
\begin{align}
D_N \equiv \frac{4d_N^2 }{N} \left(1-2^{1-\frac{4}{N}}\right) \Gamma \left( \frac{4}{N}\right) \zeta \left( \frac{4}{N}\right), \label{eq:coeffunitary}
\end{align}
where $\Gamma(x)$ and $\zeta(x)$ are gamma and zeta functions, respectively.
Equation~(\ref{eq:Tunitary}) reads $T^{5}$ for the linear point node ($N=1$) and $T^{3}$ for the quadratic point node ($N=2$). 
In the next section, we employ a similar approach for calculating $1/T_1$ for each point node.

\section{NMR relaxation rates in nonunitary point node states}
\label{sec:results}

We consider $1/T_1$ of pair potentials categorized into types (i)--(iv), assuming that the on- and off-axis point nodes do not coexist. The coexistence yields a different behavior. For instance, in the case of $(n,J_z,j_a)=(3,1,1/2)$, point nodes belonging to types (i) and (ii) coexist, and all point nodes have a linear dispersion. Thus, a point node of type (i) is not described by Eq.~(\ref{eq:generalweyl}). 
See details in Appendix~\ref{app:weyl1}. 

The temperature dependence of $1/T_1$ is summarized in Table~\ref{tab:temp_nmr}. 
The key physical quantities are the two NMR relaxation rates, $(1/T_1)_{\parallel}$ and $(1/T_1)_{\perp}$, because the two relaxation rates are identical in the unitary states. Thus, 
the difference between $(1/T_1)_{\parallel}$ and $(1/T_1)_{\perp}$ characterizes the nonunitary states.
We find that the ratio $(1/T_1)_{\parallel}/(1/T_1)_{\perp}$ can be formally described by
\begin{align}
\frac{(1/T_1)_{\parallel}}{(1/T_1)_{\perp}} \approx A' \, T^{\gamma'}, \label{eq:ratio}
\end{align}
reflects the four types of node structures. 
Coefficient $A' \in \mathbb{R}$ depends on the details of the superconducting materials. 
Meanwhile, power $\gamma'$ takes a universal value containing the Chern numbers of the point nodes as
\begin{align}
\gamma' = 
\begin{cases} 1 &\text{for type (i)}, \\ 
0 &\text{for types (ii) and (iii)}, \\
\left| \frac{2}{N} - \frac{2}{M}\right| &\text{for type (iv)}, 
\end{cases} \label{eq:diffnmr}
\end{align} 
where $N,M$ are positive integers related to Chern numbers. 
Equation~(\ref{eq:diffnmr}) includes the results in Ref.~\onlinecite{kozii2016} is a special case of type (i). 
Therefore, the temperature dependence of $(1/T_1)_{\parallel}/(1/T_1)_{\perp}$ for (i) and (iv) provides strong evidence of nonunitary states and is beneficial for understanding pairing symmetry and superconducting mechanisms.
On the other hand, $1/T_1$ for (ii) and (iii) share the same temperature dependence as the unitary chiral SCs. However, $A'$ depends on the pseudospin structure, which also provides valuable information about pairing symmetry.

In the following, we discuss the physical origin of $\gamma' \neq 0$ for (i) and (iv) in Sec.~\ref{sec:anisotropicexp} and examine how the pseudospin structure modifies $A'$ for (ii) and (iii) in Sec. ~\ref{sec:anisotropiccoeff}. 

\subsection{NMR relaxation rates for (i) and (iv)}
\label{sec:anisotropicexp}

\subsubsection{Type (i)}

First, we consider $1/T_1$ for type (i). The Weyl point nodes appear only on the rotation axis, accompanying the $\bm{q}$ vector polarized along the rotation axis. The polarized $\bm{q}$ vector gives rise to $\gamma' \neq 0$, as initially indicated in Ref.~\onlinecite{kozii2016}. Here, we generalize their results to the case of a higher Chern number $|Q_-|=2N$. 

The calculation of $1/T_1$ takes place in a similar manner to the unitary chiral SCs; however, the pseudospin anisotropically couples with the nuclear spin because of the $\bm{q}$ vector. In addition, we only consider the contribution from the $E_{{\bf k}}^-$ eigenspace, because $E_{{\bf k}}^+$ is fully gapped. 
Considering these facts, we evaluated Eq.~(\ref{eq:NMR}) in the system with $\bm{d}$ vector~(\ref{eq:generalweyl}). The energy spectrum around the point nodes is approximated as follows:
\begin{align}
E_{{\bf k}_\pm+{\bf p}}^- \simeq \sqrt{v_F^2 \hat{p}_z^2+ \tilde{v}_{\Delta}^2 \hat{p}_{\perp}^{4N} }, \label{eq:Eqgentpi}
\end{align}
and replacing the summation with the integral in terms of $E$, $1/T_1$ is recast as 
\begin{align}
\frac{1}{T_1} = &\pi \gamma_{\rm n}^2 A_{\rm hf}^2 \int_0^{\infty} D^2_-(E)\, \Big[ G(E) \frac{1+\cos^2 (\theta_I)}{2} \nonumber \\
&\qquad \qquad+ \frac{\sin^2(\theta_I)}{4} \Big]\, \left(- T \frac{\partial f_E }{\partial E} \right) dE,
\label{eq:NMRtypei}
\end{align}
where $D_-(E)$ is the density of states in $E_{\bf k_{\pm}+p}^-$ and $G(E)$ is induced by the $\bm{q}$ vector, represented as (see Appendix~\ref{app:weyl_2n})
\begin{align}
D_-(E) &= d_{2N} E^{\frac{1}{N}}, \\
G(E) &= \frac{\lambda_b^2}{4 \lambda_a^2} \frac{N'}{N} \frac{d_{2N'} }{d_{2N}} E,
\end{align}
where $N' \equiv N/(N+1)$. 
In deriving Eq.~(\ref{eq:NMRtypei}), some terms vanish via the integral of momenta on the spherical Fermi surface, and we omit the higher order terms of $\hat{p}_{\perp}$.
By performing integation, we obtain
\begin{align}
\frac{1}{T_1} = \pi \gamma_{\rm n}^2 A_{\rm hf}^2 &\Big[ \frac{1+\cos^2 (\theta_I)}{2} A_N T^{\frac{2}{N}+2} \nonumber \\ 
&+\frac{\sin^2(\theta_I)}{4} D_{2N} T^{\frac{2}{N}+1} \Big], \label{eq:NMRtypei2}
\end{align}
where
\begin{align}
A_N = &\frac{\lambda_b^2 d_{2N} d_{2N'} }{4\lambda_a^2} \frac{N'}{N}\left(\frac{1}{N}+\frac{1}{N'}\right)\nonumber \\ 
&\times \Gamma\left(\frac{1}{N}+\frac{1}{N'}\right)\zeta\left(\frac{1}{N}+\frac{1}{N'}\right) \nonumber \\ 
&\times \left( 1- 2^{1-\left(\frac{1}{N}+\frac{1}{N'}\right)}\right),
\end{align}
and $d_N$ and $D_{2N}$ are defined in Eq.~(\ref{eq:doscoeff}) and (\ref{eq:coeffunitary}) by replacing $v_{\Delta}$ with $\tilde{v}_{\Delta}$. Therefore, the origin of $\gamma' \neq 0$ in Eq.~(\ref{eq:ratio}) is the coefficient $G(E)$ in Eq.~(\ref{eq:NMRtypei}), which stems from the $\bm{q}$ vector polarized along the $z$-axis.
The results are applicable to a situation in which the Weyl point nodes appear only on the rotation axis. In Table~\ref{tab:form_factors}, $d_{\bf k}^{N=1}$ is realized for systems with $(n,J_z,j_z)=(2,1,1/2), (4,1,1/2), (6,1,1/2), (6,1,5/2)$ and $d_{\bf k}^{N=3}$ for $(6,3,3/2)$. Note that we need a fine-tuning of the parameters in $(2,1,1/2)$ and $(6,3,3/2)$ because the on-axis point nodes split into off-axis Weyl point nodes.

\subsubsection{Type (iv)}
\label{sec:NMRtypeiv}
Second, we examine $1/T_1$ for nonunitary Dirac point node states of type (iv) in a system described by the $\bm{d}$ vector~(\ref{eq:generalchiral}). 
In the evaluation of $1/T_1$, we consider low-energy excitations from $E_{\bf k}^+$ and $E_{\bf k}^-$. After following a procedure similar to that for unitary chiral SCs, Eq.~(\ref{eq:NMR}) can be rewritten as (see Appendix~\ref{app:extend})
\begin{align}
\frac{1}{T_1} = &\pi \gamma_{\rm n}^2 A_{\rm hf}^2 \int_0^{\infty} \Big\{D_+(E)D_-(E) \frac{1+\cos^2 (\theta_I)}{2} \nonumber \\
&+ [D_+^2(E)+D_-^2(E)]\frac{\sin^2(\theta_I)}{4} \Big\}\, \left(- T \frac{\partial f_E }{\partial E} \right) dE,
\label{eq:NMRtypeiv}
\end{align}
where $D_\pm(E)$ is the density of states in $E_{\bf k_{\pm}+p}^\pm$ given by
\begin{align}
D_+(E) &= d_{N}^+ E^{\frac{2}{N}}, \\
D_-(E) &= d_{M}^- E^{\frac{2}{M}}.
\end{align}
Here, $d_{N}^{\pm}$ is defined as $d_N$ [Eq.~(\ref{eq:doscoeff})], by replacing $v_{\Delta}$ with $v_{\Delta}^{\pm}$.
The $E$ dependence differs between $D_+(E)$ and $D_-(E)$, resulting in a nonzero $\gamma'$.
At low temperatures, by integrating Eq.~(\ref{eq:NMRtypeiv}) in terms of $E$ results in
\begin{align}
\frac{1}{T_1} = & \pi \gamma_{\rm n}^2 A_{\rm hf}^2 \Big[ \frac{1+\cos^2 (\theta_I)}{2} B_{N,M} T^{\frac{2}{N} +\frac{2}{M}+1} \nonumber \\
&+ \frac{\sin^2(\theta_I)}{4} \left(D_{N}^+ T^{\frac{4}{N}+1} +D_{M}^- T^{\frac{4}{M}+1}\right)\Big], \label{eq:NMRtypeiv2}
\end{align}
where
\begin{align}
B_{N,M} = &d_N^+ d_M^- \left(\frac{2}{N}+\frac{2}{M}\right) \Gamma\left(\frac{2}{N}+\frac{2}{M}\right) \zeta\left(\frac{2}{N}+\frac{2}{M}\right)\nonumber \\ 
&\times \left( 1-2^{1-\left(\frac{2}{N}+\frac{2}{M}\right)}\right),
\end{align}
and $D_N^{\pm}$ is given by Eq.~(\ref{eq:coeffunitary}) with $v_{\Delta}^{\pm}$.
The first term shows the characteristic temperature dependence due to the mixing of $D_+(E)$ and $D_-(E)$. In the second term, $D_+(E)$ and $D_-(E)$ individually contribute to the temperature dependence. When $N > M$ ($N < M$), $T^{4/N +1} $ ($T^{4/M +1} $) is the leading contribution of  $(1/T_1)_{\perp}$. Neglecting the subleading contribution, we obtain Eq.~(\ref{eq:diffnmr}). 

In Table~\ref{tab:form_factors}, the calculation above is applied to the case of $ (n,J_z,j_z)=(6,2,1/2), (6,2,5/2)$. As discussed in Sec.~\ref{sec:node_typevi}, the system realizes nonunitary Dirac point nodes with $N=1$ and $M=\pm3$. Although Eq.~(\ref{eq:dvec611}) includes $d_{\bf k}^z = \lambda_d k_z \hat{k}_+^2$, the $\lambda_d$ term only changes $v_{\Delta}^-$ to $\tilde{v}_{\Delta}^-$ within the low-energy regime. Thus, it does not affect the temperature exponent. 
Another candidate is $(n,J_z,j_z)=(6,3,1/2)$, in which nonunitary Dirac point node states with $N=2$ and $M=4$ emerge in a specific situation (see Appendix~\ref{app:6312}).

\subsection{NMR relaxation rates for (ii) and (iii)}
\label{sec:anisotropiccoeff}

Finally, we discuss the influence of the $\bm{q}$ vector on $A'$ for (ii) and (iii). 
In the following, we address nonunitary chiral SCs with the $\bm{d}$ vectors~(\ref{eq:dc2}) and (\ref{eq:dc4}) as concrete examples.

\subsubsection{Type (ii)}
\label{sec:off-axis}
\begin{figure}[tbp]
\centering
\includegraphics[width=8cm]{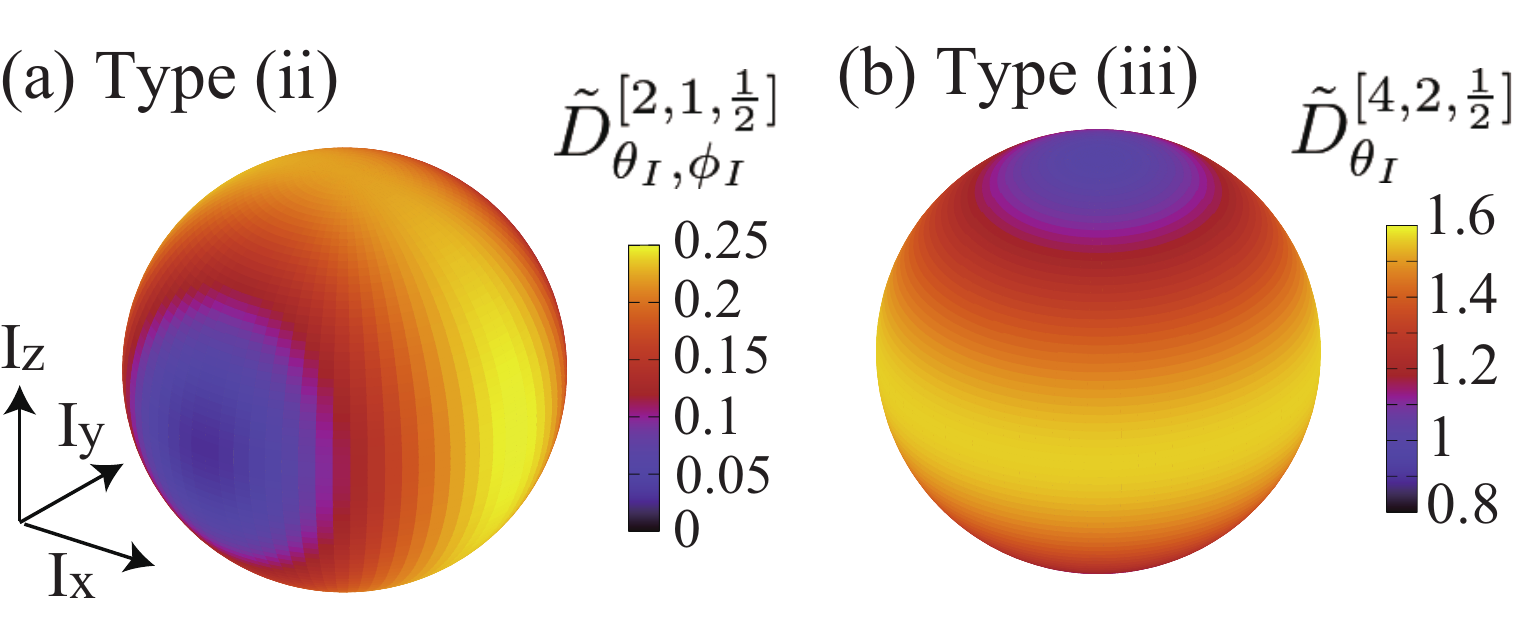}
\caption{ (Color online) Anisotropy of coefficients of $1/T_1$ in the nuclear-spin space (the direction of magnetic field), where the colors on the unit sphere represent the amplitude of the normalized coefficients: (a) $\tilde{D}^{[2,1,\frac{1}{2}]}_{\theta_I,\phi_I}=D^{[2,1,\frac{1}{2}]}_{\theta_I,\phi_I}/[7k_F^3/(270 v_F^2 v_{\Delta x}^2 v_{\Delta y}^2)]$ and (b) $\tilde{D}^{[4,2,\frac{1}{2}]}_{\theta_I}=D^{[4,2,\frac{1}{2}]}_{\theta_I}/[7k_F^3/(270 v_F^2 v_{\Delta +}^2 v_{\Delta -}^2)]$. The parameters are chosen to be (a) $\lambda_a=1, \, \lambda_b=0.5$ and (b) $\lambda_a=1, \, \lambda_b=0.5$.
}
\label{fig:coeff1}
\end{figure}

We consider a $C_2$ symmetric system using Eq.~(\ref{eq:dc2}) as an example of type (ii). From Eqs.~(\ref{eq:offnodeC2}) and (\ref{eq:enec2}), the energy spectrum around the four off-axis Weyl point nodes is given by 
\begin{align}
E_{{\bf k}_{\eta_y,\eta_z}+{\bf p} }^- \simeq \sqrt{v_F \hat{p}_z^2 + v_{\Delta x}^2 \hat{p}_x^2 + v_{\Delta y}^2 \hat{p}_y^2 }. \label{eq:enec2}
\end{align} 
Considering only the contribution from the $E_{\bf k}^-$ eigenspace, $1/T_1$ is evaluated as
\begin{align}
\frac{1}{T_1} = 4 \pi \gamma_{\rm n}^2 A_{\rm hf}^2 D^{[2,1,\frac{1}{2}]}_{\theta_I,\phi_I} \, T^5 \label{eq:NMRC2}
\end{align}
with
\begin{align}
D^{[2,1,\frac{1}{2}]}_{\theta_I,\phi_I} =& \frac{7 k_F^3}{120 v_F^2 v_{\Delta x}^2 v_{\Delta y}^2} \nonumber \\
& \times \Big[ \frac{\lambda_a \lambda_c}{(\lambda_a+\lambda_c)^2} \frac{1+\cos^2(\theta_I)}{2} \nonumber \\
& +\left( \frac{\lambda_a-\lambda_c}{\lambda_a+\lambda_c}\right)^2 \frac{\sin^2 (\theta_I)}{4} \nonumber \\
&+\frac{\lambda_a \lambda_c}{(\lambda_a+\lambda_c)^2} \frac{\cos(2\phi_I) \sin^2 (\theta_I)}{2} \Big], \label{eq:NMRC2off}
\end{align}
where we take the summation over the four off-axis Weyl point nodes (see Appendix~\ref{app:off}). The temperature dependence becomes $1/T_1 \propto T^{5}$ because the point nodes have linear dispersion. We find $\gamma' = 0$ here because the effect of the $\bm{q}$ vector is milder than that in type (i), which does not yield the prefactor $G(E)$ in Eq.~(\ref{eq:NMRtypei}). Yet, Eq.~(\ref{eq:NMRC2off}) deviates from the unitary chiral pairing states. Interestingly, Eq~(\ref{eq:NMRC2off}) follows the $C_2$ symmetry around the $z$ axis as
\begin{align}
D^{[2,1,\frac{1}{2}]}_{\theta_I,\phi_I+\pi} = D^{[2,1,\frac{1}{2}]}_{\theta_I,\phi_I}, \label{eq:coeffC2sym}
\end{align}
which reflects that the $\bm{q}$ vector preserves $C_2$ symmetry. 
In Fig.~\ref{fig:coeff1} (a), Eq. (\ref{eq:NMRC2off}) in the nuclear spin space, which is linked to real space via an applied magnetic field. We find that the coefficient retains the $C_2$ symmetry in terms of $\phi_I$. 

\subsubsection{Type (iii)}
\label{sec:extended_chiral}
We turn to the nonunitary Dirac point nodes of type (iii). 
Using Eq.~(\ref{eq:generalchiral}), $1/T_1$ is given by ~(\ref{eq:NMRtypeiv2}) with $N=M$. Thus, the temperature dependence is independent of the nuclear spin direction, which is evaluated as
\begin{align}
\frac{1}{T_1} \propto T^{\frac{4}{N}+1}.
\end{align}
Hence, the result is similar to that in Eq.~(\ref{eq:Tunitary}) in unitary Dirac point-node states. Nevertheless, the coefficients of $1/T_1$ differ from those in Eq.~(\ref{eq:Tunitary}). In the following, we discuss how the difference between $\Delta_{\bf k}^+$ and $\Delta_{\bf k}^-$ affects the coefficient of $1/T_1$.

Concretely, we consider a $C_4$ symmetric system using Eq.~(\ref{eq:dc4}), where the coefficients of $\Delta_{\bf k}^+$ and $\Delta_{\bf k}^-$ around the point nodes are different, i.e., $v_{\Delta}^+ \neq v_{\Delta}^-$. $1/T_1$ is given by Eq.~(\ref{eq:NMRtypeiv2}), satisfying $N=M=1$ as follows:
\begin{align}
\frac{1}{T_1} =\pi \gamma_{\rm n}^2 A_{\rm hf}^2 D^{[4,2,\frac{1}{2}]}_{\theta_I} \, T^5, \label{eq:NMRC4}
\end{align} 
with 
\begin{align}
D^{[4,2,\frac{1}{2}]}_{\theta_I} &= \frac{7k_F^3}{120 v_F^2(v_{\Delta}^+ v_{\Delta}^-)^2} \nonumber \\
&\times \left[\frac{1+\cos^2 (\theta_I)}{2} + \frac{(v_{\Delta}^{+})^4+(v_{\Delta}^-)^4}{(v_{\Delta}^+v_{\Delta}^-)^2} \frac{\sin^2(\theta_I)}{4}\right]. \label{eq:coeffC4}
\end{align} 
The coefficient of $1/T_1$ only depends on $\theta_I$ because the gap structure is axially symmetric. The anisotropy in terms of nuclear spin space arises from the difference between $v_{\Delta}^+$ and $v_{\Delta}^-$. In the unitary case, i.e., $v_{\Delta}=v_{\Delta}^+ = v_{\Delta}^-$, Eq.~(\ref{eq:coeffC4}) can be reduced to Eq.~(\ref{eq:Tunitary}) for $N=1$.

Equation~(\ref{eq:coeffC4}) is visualized in the nuclear spin space, that is, in the direction of the magnetic field, as shown in Fig.~\ref{fig:coeff1} (b). We found that the amplitude of the coefficient depends on $\theta_I$ in the nuclear spin space. The coefficients satisfy
\begin{align}
\left( \frac{1}{T}_1 \right)_{\parallel} \leqq \left( \frac{1}{T}_1 \right)_{\perp},
\end{align}
where $(v_{\Delta}^+)^4 + (v_{\Delta }^-)^4 \geqq 2 (v_{\Delta }^+ v_{\Delta }^-)^2$, and equality is satisfied in the unitary case. Thus, the amplitude of $1/T_1$ in the $xy$ plane increase as the difference between $v_{\Delta}^+$ and $v_{\Delta}^-$ increase. 

The above discussion is applicable when $d_{\bf k}^z=0$. This condition is satisfied in Table~\ref{tab:form_factors} except in the case of $(n,J_z,j_z)=(3,1,3/2)$. In this case, $d_{\bf k}^z$ is nonzero in the leading order, which makes the coefficient of $1/T_1$ nonaxially symmetric. See details in Appendix~\ref{app:extendc3}.

\section{Summary and discussion}
\label{sec:summary}

We developed a classification of nonunitary chiral pairing states in terms of $C_n$ symmetry and found nonunitary Dirac point nodes, from which excitations depend on the pseudospin of electrons. We categorized the nonunitary point node states into four types, (i)--(iv), in terms of the Chern number and position relative to the rotation axis. These categories cover all possible point nodes in nonunitary odd-parity chiral SCs with $C_n$ symmetry.

We examined the NMR longitudinal relaxation rate, $1/T_1$, as an observable to characterize the contribution of low-energy excitations around the point nodes in each type, and we examined the NMR longitudinal relaxation rate, $ 1/T_1 $. 
The resultant $1/T_1$ value depends on the type, as summarized in Table~\ref{tab:temp_nmr}. The deviation of $1/T_1$ from the unitary SCs is characterized by a ratio between $(1/T_1)_{\parallel}$ and $(1/T_1)_{\perp}$. The ratio is unity in unitary chiral SCs; in contrast, in nonunitary chiral SCs, the ratio satisfies the relation~(\ref{eq:ratio}) with $A'$ and $\gamma'$ being positive real numbers. We found that the ratio depends on the temperature in (i) and (iv), i.e., $\gamma' \neq 0$, which enables us to experimentally distinguish between nonunitary Dirac point, Weyl point, and unitary Dirac point nodes.
However, $1/T_1$ is independent of the temperature ($\gamma'=0$) in (ii) and (iii); namely, the temperature dependence is the same as that of the unitary chiral SCs. Nevertheless, $A'$ is not unity and reflects the configuration of the $\bm{q}$ vectors, which also contain information regarding the pairing symmetry. 

Finally, we discuss an application to nonunitary SCs in UTe$_2$. 
Recently, considerable efforts have been made to understand the heavy fermion superconductor UTe$_2$ because it is located near a ferromagnetic quantum critical point~\cite{ran2019nearly} and exhibits several anomalous properties, including a high upper critical field beyond the Pauli limit~\cite{aoki2019unconventional,nakamine2019,ran2019extreme}, reentrant superconductivity~\cite{ran2019extreme,knebel2019}, and a small reduction in the Knight shift in NMR~\cite{nakamine2019}. These findings suggest that UTe$_2$ is a potential candidate for odd-parity SCs. In addition, recent experimental studies have reported unconventional superconducting properties~\cite{aoki2022} such as the existence of point nodes~\cite{metz2019,kittaka2020}, time-reversal symmetry breaking~\cite{jiao2020chiral,hayes2021,bae2021anomalous}, and nonunitary pairing involving multiple components ~\cite{aoki2020,hayes2021,ishihara2021chiral}. Thus, UTe$_2$ is a promising platform for studying nonunitary odd-parity SCs with point nodes.

However, the symmetry of the pair potential remains to be fully elucidated, and further experimental and theoretical studies are required. From a theoretical perspective, the crystal symmetry of UTe$_2$ is $D_{2h}$; namely, possible pair potentials are classified by their irreducible representations (irreps): $A_u$, $B_{1u}$, $A_{2u}$, $A_{3u}$ for odd-parity pairing states. Furthermore, these irreps are one-dimensional, which implies that a chiral pairing state can be constructed from a mixture of two different irreps. Such a mixed state may appear under perturbation breaking $D_{2h}$ down to $C_{2h}$. For instance, when a magnetic field is applied parallel to the $z$ axis ($c$ axis in the crystal), $D_{2h}$ breaks to $C_{2h}$ with the rotation axis being the $z$ axis. Possible chiral pairing states include the $B_{1u}+iA_u$ ($C_2$-even parity) and $B_{2u} + iB_{3u}$ ($C_2$-odd parity) states, which include $8$ and $4$ Weyl point nodes~\cite{ishihara2021chiral,moriya2022} at general momenta, respectively. 

When putting these irreps into our notation, the $ B_{1u}+iA_u$ state corresponds to the case where $J_z=0$ and $B_{2u} + iB_{3u}$ states to $J_z=1$. Thus, in the former case, the pair potential is not a rotation-symmetry-protected chiral pairing state, generally leading to a fully gapped state. In fact, when the $A_u$ state is dominant, the gap structure is fully gapped as the superfluid $^3$He-B phase is. If the $B_{1u}$ state dominates the $A_u$ state, multiple point nodes appear at general momenta~\cite{ishihara2021chiral,moriya2022}. However, in the latter case, $J_z=1$, and thus, the $\bm{d}$ vector is equivalent to Eq.~(\ref{eq:dc2}). Therefore, when the Weyl point nodes are off the rotation axis, they belong to type (ii), and the associated off-axis Weyl point nodes preserve $C_2$ symmetry. When the Weyl point nodes are close to the rotation axis, they belong to type (i) and form a quadratic Weyl point node on the rotation axis. 

Our theory can be applied to $B_{2u} + iB_{3u}$ states, whose Weyl point nodes belong to either type (i) or (ii). For type (ii), $1/T_1 \propto T^{5}$ is to be expected, and the coefficient of $1/T_1$ preserves $C_2$ symmetry about the rotation axis. By contrast, for type (i), the $\bm{q}$ vector is aligned along the rotation axis, resulting in $\gamma' =1$, i.e., $(1/T_1)_{\parallel} \propto T^{4}$ and $(1/T_1)_{\perp} \propto T^{3}$.

Finally, we discuss the dependence of the Fermi surface shape. In previous theoretical studies~\cite{harima2020,fujimori2019,ishizuka2019,xu2019quasi,shick2019,shishidou2021}, ellipsoidal, cylindrical, and ring-shaped Fermi surfaces have been proposed, although the Fermi surface of UTe$_2$ is yet to be observed. Irrespective of the choice of Fermi surface, the exponent of temperature in $1/T_1$ can be detected experimentally because it only depends on the density of states and the $\bm{q}$ vector around the point node. However, the coefficient of temperature is sensitive not only to the shape of the Fermi surface, but also to other factors such as hyperfine couplings. Thus, this observation may be difficult, and more careful calculations using a realistic model may be required.

\section*{acknowledgements}
This work was supported by JSPS KAKENHI (Nos. JP18K03538, JP19H01824, JP19K14612, JP20H01857, JP21H01009, and JP22K03478), JSPS Core-to-Core Program (No. JPJSCCA20170002), and the CREST project (Grants No. JPMJCR19T2) from Japan Science and Technology Agency (JST). 

\appendix

\section{BdG Hamiltonian of nonunitary odd-parity superconductors}
\label{app:formula}
In this appendix, we summarize the properties of nonunitary odd-parity SCs. We start from the BdG Hamiltonian $H = \frac{1}{2}\sum_{\bf k}  \Psi_{\bf k} ^{\dagger} \hat{H}_{\bf k}  \Psi_{\bf k} $ with $\Psi_{\bm{k}} = (c_{{\bf k}  \uparrow},c_{{\bf k}  \downarrow},c_{ {\bf k}  \uparrow}^{\dagger},c_{ {\bf k}  \downarrow}^{\dagger})^T$ and 
\begin{equation}
 \hat{H}_{\bf k} = \begin{pmatrix} \xi_{\bf k} \hat{\bm{1}}_2 & \hat{\Delta}_{\bf k}  \\ \hat{\Delta}^{\dagger}_{\bf k}  & - \xi_{\bf k}  \hat{\bm{1}}_2 \end{pmatrix},
\end{equation}
where $\hat{\bm{1}}_N$ is a $N \times N$ identity matrix, $\xi_{\bf k} $ is a normal Hamiltonian relative to the the Fermi level, and $c_{{\bf k} s}^{\dagger}$ $(c_{{\bf k} s})$ is the creation (annihilation) operator of electron with pseudospin $s \in \{\uparrow, \downarrow \}$. $\hat{\Delta}_{\bf k}$ is a $2 \times 2$ matrix of pair potential defined by Eq.~(\ref{eq:pairpot}). 

 The BdG Hamiltonian is diagonalized by the generalized Bogoliubov transformation $\hat{U}_{\bf k}$ such that
 \begin{align}
 \hat{E}_{\bf k}= \hat{U}_{\bf k}^{\dagger} \hat{H}_{\bf k} \hat{U}_{\bf k},
 \end{align}
 with the eigenvalues and unitary operator,
 \begin{align}
 \hat{E}_{\bf k}&= \begin{pmatrix} E_{{\bf k}}^+&0&0&0 \\ 
                                             0&E_{{\bf k}}^-&0&0 \\
                                             0&0&-E_{{\bf k}}^+&0 \\
                                             0&0&0&-E_{{\bf k}}^- \end{pmatrix}, \label{eq:bdgeigenvalue}\\
 \hat{U}_{\bf k} &= \begin{pmatrix} \hat{u}_{\bf k} & \hat{v}_{\bf k} \\ \hat{v}_{-{\bf k}}^{\ast} & \hat{u}_{-{\bf k}}^{\ast} \end{pmatrix}. \label{eq:bdgeigenvector}
 \end{align}
 Here the $2 \times 2$ matrices $\hat{u}_{\bf k}$ and $\hat{v}_{\bf k}$ are defined by
 \begin{align}
  c_{\bm{k}s} = \sum_{\alpha} (\hat{u}_{{\bf k} s\alpha} a_{ {\bf k} \alpha} + \hat{v}_{{\bf k} s\alpha} a_{- {\bf k} \alpha}^{\dagger}), \label{eq:c-atrans}
 \end{align} 
 where $a_{ {\bf k} \alpha}^{\dagger}$ $(a_{ {\bf k} \alpha})$ is the creation (annihilation) operator of a quasiparticle. 
 Defining $|{\bf k} \alpha \rangle = (u_{{\bf k} \uparrow \alpha},u_{{\bf k} \downarrow \alpha}, v_{-{\bf k} \uparrow \alpha}^{\ast},v_{-{\bf k} \downarrow \alpha}^{\ast})^{\rm T}$, the eigenvalue equation is rewritten as
 \begin{align}
  \hat{H}_{\bf k}|{\bf k} \alpha \rangle= E_{\bf k}^{\alpha} |{\bf k} \alpha \rangle.
 \end{align}
  In the following, we show the explicit forms of Eqs.~(\ref{eq:bdgeigenvalue}) and (\ref{eq:bdgeigenvector}) for (non)unitary pair potentials, which satisfy $\hat{\Delta}_{\bf k} \hat{\Delta}_{\bf k}^{\dagger} = |\bm{d}_{\bf k}| \hat{\bm{1}}_2$ ($\hat{\Delta}_{\bf k} \hat{\Delta}_{\bf k}^{\dagger} = |\bm{d}_{\bf k}| \hat{\bm{1}}_2 + \bm{q}_{\bf k} \cdot \bm{\sigma}$). 
 
 For the unitary odd-parity pairing, Eqs.~(\ref{eq:bdgeigenvalue} and (\ref{eq:bdgeigenvector}) are given by Eq.~(\ref{eq:eigenuni}) and 
 \begin{align}
  &\hat{u}_{\bf k} = \hat{\bm{1}}_2 \sqrt{\frac{1}{2} \left(1+ \frac{\xi_{\bf k}}{E_{\bf k}}\right)} , \label{eq:ustate_uni} \\
  &\hat{v}_{\bf k} = \frac{-\hat{\Delta}_{\bf k}}{|\bm{d}_{\bf k}|} \sqrt{\frac{1}{2} \left(1- \frac{\xi_{\bf k}}{E_{\bf k}}\right)}. \label{eq:vstate_uni} 
 \end{align}

 On the other hand, for the nonunitary odd-parity pairings, the energy spectra are given by Eq.~(\ref{eq:enespec}).
 The associated $\hat{u}_{\bf k}$ and $\hat{v}_{\bf k}$ are described as~\cite{sigrist1991}
 \begin{widetext}
 \begin{align}
 \hat{u}_{\bf k} &= \begin{pmatrix} u_{{\bf k}\uparrow +} & u_{{\bf k}\uparrow -} \\ u_{{\bf k}\downarrow +} & u_{{\bf k}\downarrow -} \end{pmatrix} \nonumber \\
 &=\frac{1}{S_{\bf k}} \Bigg[ \sqrt{\frac{1}{2} \left(1+ \frac{\xi_{\bf k}}{E_{{\bf k}}^+}\right)} (|\bm{q}_{\bf k}| \hat{\bm{1}}_2 + \bm{q}_{\bf k} \cdot \bm{\sigma})(\hat{\bm{1}}_2 + \sigma_z) + \sqrt{\frac{1}{2} \left(1+ \frac{\xi{\bf k}}{E_{{\bf k}}^-}\right)} (|\bm{q}_{\bf k}| \hat{\bm{1}}_2 - \bm{q}_{\bf k} \cdot \bm{\sigma})(\hat{\bm{1}}_2 - \sigma_z)\Bigg], \label{eq:ustate}\\
  \hat{v}_{\bf k} &= \begin{pmatrix} v_{{\bf k}\uparrow +} & v_{{\bf k}\uparrow -} \\ v_{{\bf k}\downarrow +} & v_{{\bf k}\downarrow -} \end{pmatrix} \nonumber \\
 &=\frac{-i}{S_{\bf k}} \Bigg[ \sqrt{\frac{1}{2} \left(1- \frac{\xi_{\bf k}}{E_{{\bf k}}^+}\right)} \frac{\bm{D}_{{\bf k} }^-  \cdot \bm{\sigma} \sigma_y}{\sqrt{|\bm{d}_{\bf k}|^2+|\bm{q}_{\bf k}|}}(\hat{\bm{1}}_2 + \sigma_z) +\sqrt{\frac{1}{2} \left(1- \frac{\xi_{\bf k}}{E_{{\bf k}}^-}\right)} \frac{\bm{D}_{{\bf k} }^+ \cdot \bm{\sigma} \sigma_y}{\sqrt{|\bm{d}_{\bf k}|^2-|\bm{q}_{\bf k}|}}(\hat{\bm{1}}_2 - \sigma_z)\Bigg],  \label{eq:vstate}
 \end{align}
 \end{widetext}
 with $S_{\bf k}=\sqrt{8(|\bm{q}_{\bf k}|^2+q_z |\bm{q}_{\bf k}|)}$ and $\bm{D}_{{\bf k}}^\pm = |\bm{q}_{\bf k}| \bm{d}_{\bf k} \pm i \bm{d}_{\bf k} \times \bm{q}_{\bf k}$.
 
 \section{Other gap structures}
  \label{app:othergap}
 \subsection{Gap structure of $(3,1,1/2)$}
 \label{app:3112}
Another type of splitting of Weyl point nodes occurs in $(n,J_z,j_z)=(3,1,1/2)$, where Weyl point nodes exist on and off the rotation axis. Adding $C_3$-symmetry-preserving perturbations modify Eqs.~(\ref{eq:dc6}) and (\ref{eq:qc6}) as
\begin{align}
 \bm{d}^{[3,1,\frac{1}{2}]}_{\bf k} &= [\lambda_a \hat{k}_z +\lambda_c \hat{k}_-, i( \lambda_a \hat{k}_z -\lambda_c \hat{k}_-), \lambda_b \hat{k}_+], \label{eq:dc3} \\
 \bm{q}^{[3,1,\frac{1}{2}]}_{\bf k} &= -2\Big(\lambda_a\lambda_b \hat{k}_x \hat{k}_z-\lambda_b\lambda_c(\hat{k}_x^2-\hat{k}_y^2), \nonumber\\ 
              &  \quad \qquad \lambda_a\lambda_b \hat{k}_y \hat{k}_z+2\lambda_b\lambda_c \hat{k}_x \hat{k}_y, \nonumber \\
              &  \quad \qquad -\lambda_a^2 \hat{k}_z^2+\lambda_c^2 \hat{k}_{\perp}^2\Big). \label{eq:qc3}
\end{align}
 Under the $C_3$ symmetry, the on-axis point node at the north (south) pole with $Q_-=2(-2)$ changes to one on-axis point node with $Q_-=-1(1)$ and three off-axis point nodes with $Q_-=1(-1)$. Therefore, we have eight point nodes in total.  The position of off-axis point nodes is determined from 
 \begin{align}
  \cos(3 \phi) =\pm1, \ \ \tan(\theta) = \mp \frac{4\lambda_a \lambda_c}{\lambda_b^2}, \label{eq:offpositionC3}
 \end{align}
where the double sign corresponds.
 
 \subsection{Gap structure of $(6,3,1/2)$}
 \label{app:6312}
We consider the node structure of $(n,J_z,j_z)=(6,3,1/2)$. The $\bm{d}$ vector is represented as
\begin{align}
 \bm{d}^{[6,3,\frac{1}{2}]}_{\bf k} = &( \lambda_a \hat{k}_z \hat{k}_+^2+\lambda_b \hat{k}_z \hat{k}_-^2, \nonumber \\ 
 &i(\lambda_a \hat{k}_z \hat{k}_+^2-\lambda_b \hat{k}_z \hat{k}_-^2),\lambda_c \hat{k}_+^3 +\lambda_d \hat{k}_-^3), \label{eq:dvec631}
\end{align}
which consists only of  $k$-cubic terms. For brevity, we set $\lambda_d=0$. The gap structure is categorized into two cases depending on $\lambda_b$. First we consider the case of $\lambda_b=0$, in which nonunitary Dirac point nodes of type (iv) appear.
Around the point nodes, the dispersion relation is given by 
\begin{align}
 \Delta_{{\bf k}_{\pm}+{\bf p}}^+ \simeq 2\lambda_a \hat{p}_{\perp}^2, \ \  \Delta_{{\bf k}_{\pm}+{\bf p}}^- \simeq \frac{\lambda_c^2}{2 \lambda_a} \hat{p}_{\perp}^4,  
\end{align}
 and their Chern numbers are calculated as $Q_+=2$ and $Q_-=4$ at the north pole.
When $\lambda_b $ is turn on, the point nodes in $\Delta_{\bf k}^-$ split into six off-axis Weyl point nodes with $|Q_-|=1$ through the change of the Chern number from $Q_-=4$ to $-2+1 \times 6$, where the on-axis point remains as a nonunitary Dirac point node of type (iii) with quadratic dispersion. As a result, two nonunitary Dirac point nodes and twelve off-axis Weyl point nodes appear. The position of the off-axis Weyl point nodes is given by
\begin{align}
\cos(6\phi)=\pm1 , \ \ \tan^2 (\theta) = \mp \frac{4 \lambda_a \lambda_b}{\lambda_c^2}.
\end{align}
 \subsection{Gap structure of $(6,3,3/2)$}
 \label{app:6332}
Finally, we discuss off-axis Weyl point nodes for $(n,J_z,j_z)=(6,3,3/2)$. The $\bm{d}$ vector including $k$-cubic terms is given by
\begin{align}
  \bm{d}^{[6,3,\frac{3}{2}]}_{\bf k}= ((\lambda_a+\lambda_b) \hat{k}_z, i(\lambda_a-\lambda_b) \hat{k}_z,\lambda_c \hat{k}_+^3 +\lambda_d \hat{k}_-^3). \label{eq:dvec633}
\end{align} 
In the absence of the cubic terms, both $E_{{\bf k}+}$ and $E_{{\bf k}-}$ have a line node at $k_z=0$. Adding the $k$-cubic terms changes the line node to twelve off-axis Weyl point nodes.
The position of off-axis Weyl point nodes is given by
 \begin{align}
\cos(6\phi)=\pm1 , \ \ \frac{\hat{k}_{\perp}^6}{\hat{k}_z^2}  = \mp \frac{4 \lambda_a \lambda_b}{\lambda_c^2},
\end{align}
where we put $\lambda_d=0$.
Note that, when we choose $\lambda_a=0$ or $\lambda_b=0$, the off-axis Weyl point nodes meet at the rotation axis and create a Weyl point node of type (i) with $|Q_-|=6$.  
 
 \section{Calculations of NMR relaxation rates }
 \label{app:calculation}
 In this appendix, we show the derivation of $1/T_1$ in the main paragraph.  
 For the evaluation of $1/T_1$, Eq~(\ref{eq:NMR}) is rewritten, using the generalized Bogoliubov transformation [Eq.~(\ref{eq:c-atrans})] and $\langle {\bf k}_1 \alpha |a^{\dagger}_{{\bf k}_2 \alpha'}a_{{\bf k}_3 \beta'}|{\bf k}_4 \beta\rangle = \delta_{{\bf k}_1,{\bf k}_2}\delta_{{\bf k}_3,{\bf k}_4} \delta_{\alpha, \alpha'} \delta_{\beta,\beta'}$, as
\begin{align}
\frac{1}{T_1} = &2 \pi \gamma_n^2 A_{\rm hf}^2 \sum_{\alpha,\beta}\sum_{\bf k,k'} \nonumber \\
&\times \Big|\langle -\bm{I}| I_+ |\bm{I} \rangle (u_{{\bf k} \downarrow \alpha}^{\ast}u_{{\bf k}' \uparrow \beta}-v_{-{\bf k}' \downarrow \beta}^{\ast}v_{-{\bf k} \uparrow \alpha}) \nonumber \\
&+\langle -\bm{I}| I_- |\bm{I} \rangle (u_{{\bf k} \uparrow \alpha}^{\ast}u_{{\bf k}' \downarrow \beta}-v_{-{\bf k}' \uparrow \beta}^{\ast}v_{-{\bf k} \downarrow \alpha})  \nonumber \\
&+\langle -\bm{I}| I_z |\bm{I} \rangle (u_{{\bf k} \uparrow \alpha}^{\ast}u_{{\bf k}' \uparrow \beta}-v_{-{\bf k}' \uparrow \beta}^{\ast}v_{-{\bf k} \uparrow \alpha} \nonumber \\
& \qquad \qquad \quad-u_{{\bf k} \downarrow \alpha}^{\ast}u_{{\bf k}' \downarrow \beta}+v_{-{\bf k}' \downarrow \beta}^{\ast}v_{-{\bf k} \downarrow \alpha})\Big|^2 \nonumber \\
&\times f_{{\bf k}'}^\beta (1-f_{\bf k}^\alpha)  \delta(E_{\bf k}^\alpha - E_{{\bf k}'}^\beta), \label{eq:NMRexpand}
\end{align}
where we define $I_{\pm} = (I_x\pm i I_y)/2$ and neglect a pair excitation process. From Eq.~(\ref{eq:eigen_nuclear}), the expectation values of the nuclear spin are calculated  as
\begin{subequations} \label{eq:exp_nuclear}
\begin{align}
\langle -\bm{I}| I_+ |\bm{I} \rangle &= e^{i \phi_I} \sin^2(\theta_I), \\
\langle -\bm{I}| I_- |\bm{I} \rangle &= -e^{-i \phi_I} \cos^2(\theta_I), \\
\langle -\bm{I}| I_z |\bm{I} \rangle &= \frac{\sin(\theta_I)}{2}.
\end{align}
\end{subequations}
 In addition, $u_{{\bf k} s \alpha}$ and $v_{{\bf k} s \alpha}$ are given by  Eqs.~(\ref{eq:ustate_uni}) and (\ref{eq:vstate_uni}) [Eqs.~(\ref{eq:ustate}) and (\ref{eq:vstate})] for the (non-)unitary pair potentials. Using this, we proceed with the calculation of $1/T_1$ in the low-energy regime $T \ll \Delta_0$. Hereafter, we assume a spherical Fermi surface for simplicity.
 
  \subsection{Unitary}
  \label{app:nmruni}
  To begin with, we consider $1/T_1$ of the unitary state. Rewriting Eqs.~(\ref{eq:ustate_uni}) and (\ref{eq:vstate_uni}) as $\hat{u}_{\bf k} = u_{\bf k}\hat{\bm{1}}_2$ and $\hat{v}_{\bf k} = v_{\bf k} \hat{\Delta}_{\bf k} $, Eq.~(\ref{eq:NMR}) is recast as
  \begin{align}
  \frac{1}{T_1} = &2 \pi \gamma_n^2 A_{\rm hf}^2 \sum_{\alpha,\beta}\sum_{\bf k,k'} \nonumber \\
  &\times \Big|\sum_i \sum_{ss'} \langle -\bm{I}| I_i |\bm{I} \rangle (\sigma_i)_{ss'} (u_{\bf k}^{\ast}u_{\bf k'} \delta_{s \alpha} \delta_{s' \beta} \nonumber \\
  &\qquad - v_{\bf k'}^{\ast} v_{\bf k} (\Delta_{\bf k'}^{\ast})_{s \beta} (\Delta_{\bf k})_{s' \alpha}) \Big|^2 \nonumber \\
  &\times f_{{\bf k}'} (1-f_{\bf k})  \delta(E_{\bf k} - E_{{\bf k}'}) \nonumber \\
  =& 2 \pi \gamma_n^2 A_{\rm hf}^2 \sum_{\bf k,k'} \nonumber \\
  &\times \Big\{ \left( |u_{\bf k}|^2|u_{\bf k'}|^2 + |\Delta_{\bf k'}|^2|\Delta_{\bf k}|^2|v_{\bf k'}|^2|v_{\bf k}|^2\right) \nonumber \\
  & - \sum_{ij}\langle -\bm{I}| I_i |\bm{I} \rangle (\langle -\bm{I}| I_j |\bm{I} \rangle)^{\ast} \nonumber \\
  & \times  \Big[\tr(\sigma_i \Delta_{\bf k'}^{\rm T} \sigma_j^{\ast} \Delta_{\bf k})u_{\bf k}^{\ast}u_{\bf k'} v_{\bf k'}v_{\bf k}^{\ast} \nonumber \\
  & \qquad +\tr(\Delta_{\bf k'}^{\dagger} \sigma_i\Delta_{\bf k} \sigma_j^{\ast} )u_{\bf k}u_{\bf k'}^{\ast} v_{\bf k'}^{\ast} v_{\bf k}\Big]\Big\} \nonumber \\
  &\times f_{{\bf k}'} (1-f_{\bf k})  \delta(E_{\bf k} - E_{{\bf k}'}) , \label{eq:NMRexpand_uni}
  \end{align}
  where we use $\tr(\sigma_i \sigma_j ) = 2 \delta_{ij}$ and $\sum_{i} |\langle -\bm{I}|I_i|\bm{I}\rangle |^2 = 1/2$. The last terms describe the coherence factor, which remains nonzero for $s$-wave SCs. In contrast, in chiral SCs, this term vanishes under the integral over momenta on the Fermi surface. Substituting Eqs.~(\ref{eq:ustate_uni}) and (\ref{eq:vstate_uni}) into Eq.~(\ref{eq:NMRexpand_uni}) yields
    \begin{align}
  \frac{1}{T_1} =& \pi \gamma_n^2 A_{\rm hf}^2 \sum_{\bf k,k'} \left(1+\frac{\xi_{\bf k} \xi_{\bf k'}}{E_{\bf k}E_{\bf k'}}\right) \nonumber \\
   &\times f_{{\bf k}'} (1-f_{\bf k})  \delta(E_{\bf k} - E_{{\bf k}'}) ,
  \end{align}
  where the second term vanishes under the integral of momenta. Thus, we arrive at Eq.~(\ref{eq:NMRunitary}).
  
 \subsection{Type (i)}
 \label{app:weyl_2n}
 We now turn to the nonunitary cases. We start from the derivation of Eq.~(\ref{eq:NMRtypei}). The $\bm{d}$ vector is given by Eqs.~(\ref{eq:generalweyl}). Since $E_{\bf k}^{+}$ is fully gapped, we only consider Eq.~(\ref{eq:NMRexpand}) with  $\alpha = \beta =-$. Substituting the $\bm{d}$ vector into Eqs.~(\ref{eq:ustate}) and (\ref{eq:vstate}) and expanding them around the point nodes to the leading order of ${\bf p}$, low-energy effective forms are given by
\begin{subequations} \label{eq:uv_generalWeyln}
\begin{align}
u_{{\bf p}\uparrow -}  &= \frac{\lambda_b \hat{p}_-^N}{2\lambda_a} \sqrt{\frac{1}{2}\left(1+ \frac{\xi_{\bf p}}{E_{{\bf p}}^-}\right)}, \\
u_{{\bf p}\downarrow -}  &= \sqrt{\frac{1}{2}\left(1+ \frac{\xi_{\bf p}}{E_{{\bf p}}^-}\right)}, \\
v_{{\bf p}\uparrow -} &= -\frac{\lambda_b \hat{p}_+^N}{2\lambda_a} \sqrt{\frac{1}{2}\left(1- \frac{\xi_{\bf p}}{E_{{\bf p}}^-}\right)},  \\
v_{{\bf p}\downarrow -}& = -\frac{\hat{p}_+^{2N}}{\hat{p}_{\perp}^{2N}} \sqrt{\frac{1}{2}\left(1- \frac{\xi_{\bf p}}{E_{{\bf p}}^-}\right)},
\end{align}   
\end{subequations}
where $\xi_{\bf p} = v_F p_z$. We relabel the subscript as $E_{\bf p+k_{\pm}} \to E_{\bf p}$ for the simplicity of notation.
Then, $1/T_1$ is calculated by substituting Eqs.~(\ref{eq:uv_generalWeyln}) and (\ref{eq:exp_nuclear}) into Eq.~(\ref{eq:NMRexpand}) and replacing the summation with the integral. Keeping the leading contributions of ${\bf p}$, Eq.~(\ref{eq:NMRexpand}) is recast as
 \begin{align}
  \frac{1}{T_1} = &2 \pi   \gamma_{\rm n}^2 A_{\rm hf}^2 \frac{1}{(2\pi)^6} \int d^3 p \int d^3 p' \nonumber \\
 &\times  \Big[ \frac{1+\cos^2 (\theta_I)}{2} g^{1N}_{\bf p,p'} + \frac{\sin^2(\theta_I)}{4}  g^{2N}_{\bf p,p'} \Big] \nonumber \\
 &\times f_{{\bf p}'}^- (1-f_{\bf p}^-) \delta(E_{{\bf p}}^- - E_{{\bf p}'}^-), \label{eq:NMRtypeiapp}
 \end{align}
 where  $g^{1N}_{\bf p,p'}$ and $g^{2N}_{\bf p,p'}$ are given by
\begin{align}
&g^{1N}_{\bf p,p'} = \frac{\lambda_b^2 \hat{p}_{\perp}^N}{8\lambda_a^2} \left(1+ \frac{\xi_{{\bf p} }\xi_{{\bf p}' }}{E_{{\bf p} }^-E_{{\bf p}' }^-}\right), \label{eq:generalCF1}\\
&g^{2N}_{\bf p,p'} = \frac{1}{2}\left(1+ \frac{\xi_{{\bf q} }\xi_{{\bf q}' }}{E_{{\bf p} }^-E_{{\bf p}' }^-}\right). \label{eq:generalCF2}
\end{align}
When integrating ${\bf p}$ on the Fermi surface, the second term of Eqs.~(\ref{eq:generalCF1}) and (\ref{eq:generalCF2}) vanishes. Thus, using the density of states, Eq.~(\ref{eq:NMRtypeiapp}) is rewritten as Eq.~(\ref{eq:NMRtypei}). Performing the integral in terms of $E$, we finally get Eq.~(\ref{eq:NMRtypei2}). 
 
 \subsection{Type (ii)}
 \label{app:off}
 Next, we show the derivation of Eq.~(\ref{eq:NMRC2}). Similarly to Sec.~\ref{app:weyl_2n}, the main contribution to $1/T_1$ comes from the $E_{{\bf k}-}$ eigenspace. The $u_{{\bf k} s -}$ and $v_{{\bf k} s -}$ expanded around the Weyl point nodes to the leading order of ${\bf p} ={\bf k}- {\bf k}_{\eta_y,\eta_z}$ become
 \begin{subequations} \label{eq:uvC2off}
\begin{align}
u_{{\bf p}\uparrow -}  = &\frac{-i \eta_y \sqrt{\lambda_c}}{\sqrt{\lambda_a+\lambda_c}} \sqrt{\frac{1}{2}\left(1+ \frac{\xi_{\bf p}}{E_{\bf p}^-}\right)}, \\
u_{{\bf p}\downarrow -}  = &\frac{\sqrt{\lambda_a}}{\sqrt{\lambda_a+\lambda_c}} \sqrt{\frac{1}{2}\left(1+ \frac{\xi_{\bf p}}{E_{{\bf p}}^-}\right)}, \\
v_{{\bf p}\uparrow -} = &\frac{ \sqrt{\lambda_a}}{\sqrt{\lambda_a+\lambda_c}} \frac{\eta_z \lambda_b \hat{p}_x + i \sqrt{\lambda_b^2+4 \lambda_a \lambda_c} \hat{p}_y}{\sqrt{\lambda_b^2 \hat{p}_x^2+(\lambda_b^2+4 \lambda_a \lambda_c) \hat{p}_y^2}} \nonumber \\
&\times \sqrt{\frac{1}{2}\left(1- \frac{\xi_{\bf q}}{E_{{\bf p}}^-}\right)},  \\
v_{{\bf p}\downarrow -} = &\frac{-i \eta_y \sqrt{\lambda_c}}{\sqrt{\lambda_a+\lambda_c}} \frac{\eta_z \lambda_b \hat{p}_x + i \sqrt{\lambda_b^2+4 \lambda_a \lambda_c} \hat{p}_y}{\sqrt{\lambda_b^2 \hat{p}_x^2+(\lambda_b^2+4 \lambda_a \lambda_c) \hat{p}_y^2}} \nonumber \\
&\times \sqrt{\frac{1}{2}\left(1- \frac{\xi_{\bf p}}{E_{{\bf p}}^-}\right)},
\end{align}   
\end{subequations}
where we rotate {\bf p} as ${\bf p} \to (p_x,p_y \cos(\theta_{\eta_y,\eta_z})+p_z \sin(\theta_{\eta_y,\eta_z}), -p_y \sin(\theta_{\eta_y,\eta_z})+p_z \cos(\theta_{\eta_y,\eta_z}))$ with
\begin{subequations}
\begin{align}
\cos(\theta_{\eta_y,\eta_z}) &= \frac{\eta_z \lambda_b}{\sqrt{\lambda_b^2+4\lambda_a \lambda_c}}, \\ 
\sin(\theta_{\eta_y,\eta_z}) &= \frac{2 \eta_y \sqrt{\lambda_a\lambda_c} }{\sqrt{\lambda_b^2+4\lambda_a \lambda_c}}, 
\end{align}
\end{subequations}
to eliminate the $p_z$ dependence.
Substituting the eigenstates to Eq.~(\ref{eq:NMRexpand}), we obtain:
\begin{align}
\frac{1}{T_1} = &2 \pi  \gamma_{\rm n}^2 A_{\rm hf}^2 \frac{1}{(2\pi)^6} \int d^3 p \int d^3 p' \nonumber \\
&\times g_{\rm off} (\theta_I,\phi_I) \frac{1}{2}  \left(1+ \frac{\xi_{\bf p}\xi_{\bf p'}}{E_{{\bf p}}^-E_{{\bf p'}}^-}\right) \nonumber \\
 &\times f_{{\bf p}'}^- (1-f_{\bf p}^-) \delta(E_{{\bf p}}^- - E_{{\bf p}'}^-), \label{eq:NMRC2pre}
\end{align}
with
\begin{align}
 g_{\rm off}(\theta_I,\phi_I) = &\frac{\lambda_a \lambda_c}{(\lambda_a+\lambda_c)^2} \frac{1+\cos^2(\theta_I)}{2}  \nonumber \\
& +\left( \frac{\lambda_a-\lambda_c}{\lambda_a+\lambda_c}\right)^2 \frac{\sin^2 (\theta_I)}{4}  \nonumber \\
&+\frac{\lambda_a \lambda_c}{(\lambda_a+\lambda_c)^2} \frac{\cos(2\phi_I) \sin^2 (\theta_I)}{2} \nonumber \\
&-\eta_y \frac{\sqrt{\lambda_a\lambda_c}(\lambda_a-\lambda_c)}{(\lambda_a+\lambda_c)^2} \sin (\theta_I) \sin(2 \theta_I) ,
\end{align}
where the last term depends on $\eta_y$, so that it vanishes under the summation of all point nodes. After integrating Eq.~(\ref{eq:NMRC2pre}) in terms of ${\bf p}, {\bf p}'$, we get Eq.~(\ref{eq:NMRC2}).

\subsection{Types (iii) and (iv)}
\label{app:extend}
We can calculate $1/T_1$ for types (iii) and (iv) in a system where the $\bm{d}$ vector and energy spectrum around the point node are given by Eqs.~(\ref{eq:generalchiral}) and (\ref{eq:gaptypeiv}): type (iii) corresponds to the case with $|N|=|M|$ and type (iv) to the case with $|N|\neq |M|$. Since both $E_{{\bf k}}^+$ and $E_{{\bf k}}^-$ have the point nodes and $d_{{\bf k}z}=0$, only $u_{{\bf p} \uparrow+}$, $v_{{\bf p} \uparrow+}$, $u_{{\bf p} \downarrow-}$, and $v_{{\bf p} \downarrow-}$ are nonzero. They are given by
\begin{subequations} \label{eq:uv_extend}
\begin{align}
u_{{\bf p}\uparrow +}  &=  \sqrt{\frac{1}{2}\left(1+ \frac{\xi_{\bf p}}{E_{{\bf p}}^+}\right)}, \\
u_{{\bf p}\downarrow -}  &= \sqrt{\frac{1}{2}\left(1+ \frac{\xi_{\bf p}}{E_{{\bf p}}^-}\right)}, \\
v_{{\bf p}\uparrow +} &= \frac{\hat{p}_+^N}{\hat{p}_{\perp}^N} \sqrt{\frac{1}{2}\left(1- \frac{\xi_{\bf p}}{E_{{\bf p}}^+}\right)},  \\
v_{{\bf p}\downarrow -}& = - \frac{\hat{p}_+^M}{\hat{p}_{\perp}^M}  \sqrt{\frac{1}{2}\left(1- \frac{\xi_{\bf p}}{E_{{\bf p}}^-}\right)},
\end{align}
\end{subequations} 

Substituting Eq.~(\ref{eq:uv_extend}) into Eq.~(\ref{eq:NMRexpand}) and replacing the summation with the integral in terms of ${\bf p}, {\bf p}'$, we obtain
\begin{align}
 \frac{1}{T_1} = &2 \pi  \gamma_{\rm n}^2 A_{\rm hf}^2 \frac{1}{(2\pi)^6} \int d^3 p \int d^3 p' \nonumber \\
 &\times\Big[ \frac{\sin^2(\theta_I)}{4} \left( |u_{{\bf q}\uparrow +} |^2 |u_{{\bf p}'\uparrow +} |^2 +|v_{-{\bf p}'\uparrow +} |^2 |v_{-{\bf p}\uparrow +} |^2\right) \nonumber \\
 &\qquad \times f_{{\bf p}'}^+ (1-f_{\bf p}^+) \delta(E_{{\bf p}}^+ - E_{{\bf p}'}^+) \nonumber \\
 &+\frac{\sin^2(\theta_I)}{4} \left( |u_{{\bf p}\downarrow -} |^2 |u_{{\bf p}'\downarrow -} |^2 +|v_{-{\bf p}'\downarrow -} |^2 |v_{-{\bf p}\downarrow -} |^2\right) \nonumber \\
 &\qquad  \times f_{{\bf p}'}^- (1-f_{\bf p}^-) \delta(E_{{\bf p}}^- - E_{{\bf p}'}^-) \nonumber \\
 &+\frac{1+\cos^2(\theta_I)}{2} |u_{{\bf p}\downarrow -} |^2 |u_{{\bf p}'\uparrow +} |^2 \nonumber \\
& \qquad \times f_{{\bf p}'}^+ (1-f_{\bf p}^-) \delta(E_{{\bf p}}^- - E_{{\bf p}'}^+)  \nonumber \\
 & +\frac{1+\cos^2(\theta_I)}{2} |v_{-{\bf p}'\downarrow -} |^2 |v_{-{\bf p}\uparrow +} |^2 \nonumber \\
& \qquad \times f_{{\bf p}'}^- (1-f_{\bf p}^+) \delta(E_{{\bf p}}^+ - E_{{\bf p}'}^-)  \Big]. \label{eq:NMRtypeivapp}
\end{align}
 Since the first (second) line in Eq.~(\ref{eq:NMRtypeivapp}) only comprises $u_{{\bf p} \uparrow+}$ and $v_{{\bf p} \uparrow+}$ ($u_{{\bf p} \downarrow-}$ and $v_{{\bf p} \downarrow-}$), the integral is replaced with $\int_0^{\infty} D_+^2(E) dE$ ($\int_0^{\infty} D_-^2(E) dE$). On the other hand, the last two lines include the both contribution, so that the integral is replaced with  $\int_0^{\infty} D_+(E)D_-(E) dE$.  As a result, we obtain Eqs.~(\ref{eq:NMRtypeiv}) and (\ref{eq:NMRtypeiv2})

\subsection{$1/T_1$ for $(3,1,1/2)$}
\label{app:weyl1}
The gap structure of $(n,J_z,j_z)=(3,1,1/2)$ consists of on-and off-axis Weyl point nodes, which induce a slightly different $1/T_1$. For off-axis Weyl point nodes, there are six point nodes with $|Q_-|=1$ [Eq.~(\ref{eq:offpositionC3})].  The calculation is similar to the case of $(2,1,1/2)$. When we take into account the contribution from the six off-axis Weyl point nodes, $1/T_1$ yields
\begin{align}
 \frac{1}{T_1} =  6 \pi \gamma_{\rm n}^2 A_{\rm hf}^2  D^{[3,1,\frac{1}{2}]}_{{\rm off}, \theta_I} \, T^5 \label{eq:NMRC3off}
\end{align}
with
\begin{align}
D^{[3,1,\frac{1}{2}]}_{{\rm off}, \theta_I} = & \frac{7k_F^3}{120 v_F^2 v_{\Delta x}^2 v_{\Delta y}^2} \nonumber \\
          &\times \Big[ \frac{4 \lambda_b^2 \lambda_c^2}{(\lambda_b^2+4\lambda_c^2)^2} \frac{1+\cos^2(\theta_I)}{2} \nonumber \\
          &+\left( \frac{\lambda_b^2-4\lambda_c^2}{\lambda_b^2+4 \lambda_c^2}\right)^2 \frac{\sin^2 (\theta_I)}{4}  \Big],  
\end{align}
where the energy spectrum around the off-axis point nodes is given by 
\begin{align}
E_{{\bf p}}^- = \sqrt{v_F^2 \hat{p}_z^2 + \tilde{v}_{\Delta x}^{ 2} \hat{p}_x^2+ \tilde{v}_{\Delta y}^{ 2} \hat{p}_y^2}
\end{align}
with $\tilde{v}_{\Delta x}=2\lambda_c \sqrt{\lambda_b^4+16 \lambda_a^2 \lambda_c^2}/(\lambda_b^2+4 \lambda_c^2)$ and $\tilde{v}_{\Delta y}=6 \lambda_b^2 \lambda_c /(\lambda_b^2+4 \lambda_c^2)$. Vanishing of $\phi_I$ dependent terms in Eq.~(\ref{eq:NMRC3off}) stems from the $C_3$ symmetry.

On the other hand, the on-axis Weyl point nodes read a different temperature dependence due to the $\bm{q}$ vector polarized along the $z$ axis.  The eigenvalues and eigenstates expanded around the point node to the leading order of ${\bf p} = {\bf k} -{\bf k}_{\pm}$ are given by
 \begin{align}
 E_{{\bf p}}^- = \sqrt{v_F^2 \hat{p}_z^2+ v_{\Delta}^2 \hat{q}_{\perp}^2 }, \label{eq:eneQ1}
 \end{align}
and
\begin{subequations} \label{eq:uv_weyl1}
\begin{align}
u_{{\bf p}\uparrow -}  &= \frac{\lambda_b \hat{p}_-}{2\lambda_a} \sqrt{\frac{1}{2}\left(1+ \frac{\xi_{\bf p}}{E_{{\bf p}}^-}\right)}, \\
u_{{\bf p}\downarrow -}  &= \sqrt{\frac{1}{2}\left(1+ \frac{\xi_{\bf p}}{E_{{\bf p}}^-}\right)}, \\
v_{{\bf p}\uparrow -} &= -i\frac{\lambda_b \hat{p}_+^2}{2\lambda_a \hat{p}_{\perp}} \sqrt{\frac{1}{2}\left(1- \frac{\xi_{\bf p}}{E_{{\bf p}}^-}\right)},  \\
v_{{\bf p}\downarrow -}& = -i\frac{\hat{p}_-}{\hat{p}_{\perp}} \sqrt{\frac{1}{2}\left(1- \frac{\xi_{\bf p}}{E_{{\bf p}}^-}\right)},
\end{align}   
\end{subequations}
where $v_{\Delta}=2|\lambda_c|$. Thus,  the low-energy forms of $u_{{\bf p} s -}$ and $v_{{\bf p} s -}$ are similar to Eq.~(\ref{eq:uv_generalWeyln}), while the energy spectrum change from the quadratic dispersion to the linear dispersion.  Substituting Eqs.~(\ref{eq:eneQ1}) and (\ref{eq:uv_weyl1}) into Eq.~(\ref{eq:NMRexpand}), we obtain
\begin{align}
 \frac{1}{T_1} =  \pi \gamma_{\rm n}^2 A_{\rm hf}^2\Big[ &\frac{1+\cos^2 (\theta_I)}{2} A^{[3,1,\frac{1}{2}]}_{{\rm on}} \, T^7 \nonumber \\
 &+ \frac{\sin^2(\theta_I)}{4} B^{[3,1,\frac{1}{2}]}_{{\rm on}} \, T^5 \Big] \label{eq:NMRC3on}
\end{align} 
where
\begin{align}
 &A^{[3,1,\frac{1}{2}]}_{{\rm on}}= \frac{93\pi^2 \lambda_b^2 k_F^3}{1512 v_F^2 v_{\Delta}^6  \lambda_a^2} , \\
 &B^{[3,1,\frac{1}{2}]}_{{\rm on}}= \frac{7k_F^3}{120 v_F^2 v_{\Delta}^4 } .
\end{align}
 The temperature dependence reads as $ (1/T_1)_{\parallel} \propto T^{7}$ and $ (1/T_1)_{\perp} \propto  T^{5}$. Thus, we find $\gamma' =2$ in Eq.~(\ref{eq:ratio}), which is different from Eq.~(\ref{eq:diffnmr}). The difference comes from the linear dispersion around the point node. 
 
 In the coexistence of the on-axis and off-axis Weyl point nodes, however, the contribution from the off-axis Weyl point nodes becomes dominant. Thus, Eq.~(\ref{eq:NMRC3on}) can be masked by Eq.~(\ref{eq:NMRC3off}). Accordingly, $1/T_1 \propto T^5$ is to be expected for any direction of magnetic fields. 

\subsection{$1/T_1$ for $(3,1,3/2)$}
\label{app:extendc3}

\begin{figure}[tbp]
\centering
 \includegraphics[width=4cm]{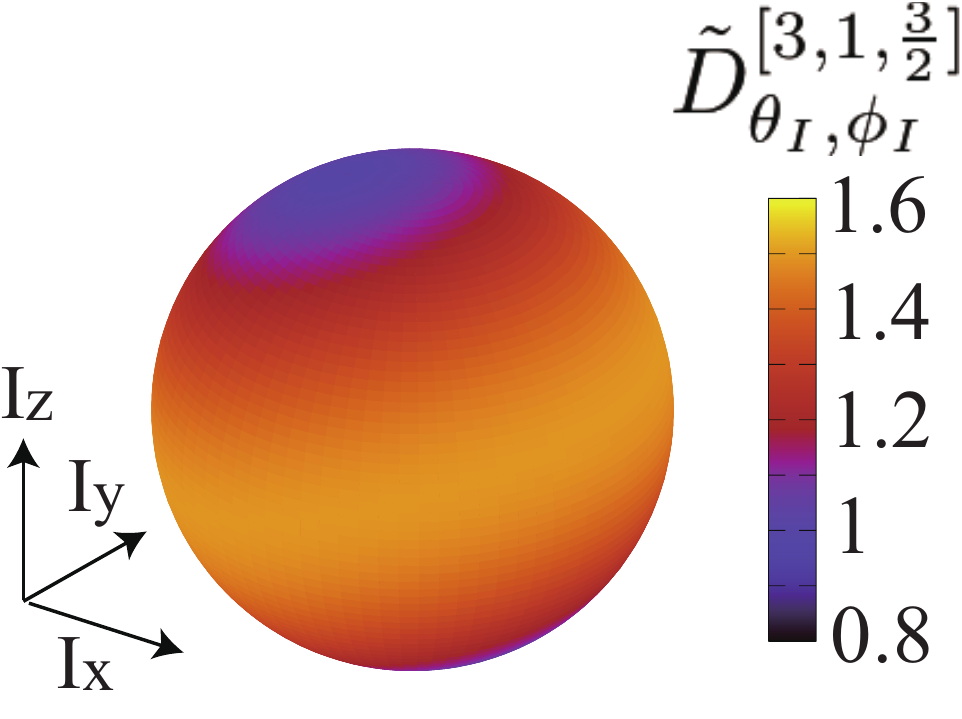}
 \caption{ (Color online) Anisotropy of the normalized coefficient in Eq.~(\ref{eq:NMR3132}). Here, the normalization is defined in such a way that $\tilde{D}_{\theta_I,\phi_I}^{[3,1,\frac{3}{2}]}=1$ in the unitary limit. The parameters are chosen to be $\lambda_a=1, \, \lambda_b=0.5 \, \lambda_c=0.5$.
 }
\label{fig:coeff2}
\end{figure}

 A special type of nonunitary Dirac point nodes of type (iii) appears in the case of $(n,J_z,j_z)=(3,1,3/2)$. The $\bm{d}$ vector and $\bm{q}$ vector are described by
  \begin{align}
 \bm{d}^{[3,1,\frac{3}{2}]}_{\bf k} &= (\lambda_a \hat{k}_+ +\lambda_b \hat{k}_+, i \lambda_a \hat{k}_+-i \lambda_b \hat{k}_+, \lambda_c \hat{k}_+), \label{eq:dc3v2} \\
 \bm{q}^{[3,1,\frac{3}{2}]}_{\bf k} &= (-2(\lambda_a-\lambda_b) \lambda_c \hat{k}_{\perp}^2,0,2(\lambda_a^2-\lambda_b^2)\hat{k}_{\perp}^2), \label{eq:qc3v2}
 \end{align}
 since the $C_3$ symmetry allows the additional term proportional to $\lambda_c$, which rotate the $\bm{q}$ vector in the $xz$ plane. The Chern number takes $Q_+=Q_-=1(-1)$ at the north (south) pole.  It should be noted that Eq.~(\ref{eq:qc3v2}) preserves the $C_3$ symmetry since the rotation operator in the pseudospin space is proportional to an identity matrix.  Interestingly, the additional term makes the pseudospin structure nonaxially symmetric, which causes a nonaxially symmetric pseudospin response.

To evaluate $1/T_1$, we approximate $u_{{\bf p}s \alpha}$ and $v_{{\bf p}s \alpha}$ around the point nodes up to the leading contribution of ${\bf p}$, which are summarized as
\begin{subequations} \label{eq:uv_special}
\begin{align}
u_{{\bf p}\uparrow +}  &=  \frac{\sqrt{f_x^2+f_z^2}+f_z}{K}\sqrt{\frac{1}{2}\left(1+ \frac{\xi_{\bf p}}{E_{{\bf p}}^+}\right)}, \\
u_{{\bf p}\downarrow +}  &= \frac{f_x}{K}\sqrt{\frac{1}{2}\left(1+ \frac{\xi_{\bf p}}{E_{{\bf p}}^+}\right)}, \\
u_{{\bf p}\uparrow -}  &=  -\frac{f_x}{K}\sqrt{\frac{1}{2}\left(1+ \frac{\xi_{\bf p}}{E_{{\bf p}}^-}\right)}, \\
u_{{\bf p}\downarrow -}  &=  \frac{\sqrt{f_x^2+f_z^2}+f_z}{K} \sqrt{\frac{1}{2}\left(1+ \frac{\xi_{\bf p}}{E_{{\bf p}}^-}\right)}, \\
v_{{\bf p}\uparrow +} &= \frac{(g_x^- +g_y^-) \hat{p}_+}{Kv_{\Delta +} \hat{p}_{\perp}} \sqrt{\frac{1}{2}\left(1- \frac{\xi_{\bf p}}{E_{{\bf p}}^+}\right)},  \\
v_{{\bf p}\downarrow +}& = - \frac{g_z^- \hat{p}_+}{Qv_{\Delta +} \hat{p}_{\perp}}  \sqrt{\frac{1}{2}\left(1- \frac{\xi_{\bf p}}{E_{{\bf p}}^+}\right)},\\
v_{{\bf p}\uparrow -} &=  \frac{g_z^+ \hat{p}_+}{Qv_{\Delta -} \hat{p}_{\perp}}  \sqrt{\frac{1}{2}\left(1- \frac{\xi_{\bf p}}{E_{{\bf p}}^-}\right)},  \\
v_{{\bf p}\downarrow +}& =  \frac{(g_x^- -g_y^-)\hat{p}_+}{Qv_{\Delta -} \hat{p}_{\perp}}  \sqrt{\frac{1}{2}\left(1- \frac{\xi_{\bf p}}{E_{{\bf p}}^-}\right)},
\end{align}
\end{subequations} 
where the coefficients are defined as
\begin{subequations} \label{eq:pre_special}
\begin{align}
f_x =& -2(\lambda_a-\lambda_b) \lambda_c, \\
f_z =& 2(\lambda_a^2-\lambda_b^2), \\
g_x^{\pm} =& 2(\lambda_a+\lambda_b)\big[(\lambda_a-\lambda_b)^2 \nonumber \\
& \qquad \qquad \mp (\lambda_a-\lambda_b)\sqrt{(\lambda_a+\lambda_b)^2+\lambda_c^2}\big] \\
g_y^{\pm} =& 2(\lambda_a+\lambda_b)\big[(\lambda_a+\lambda_b)^2+\lambda_c^2 \nonumber \\
& \qquad \qquad \mp (\lambda_a-\lambda_b)\sqrt{(\lambda_a+\lambda_b)^2+\lambda_c^2}\big] \\
g_z^{\pm} =& 2\lambda_c\big[(\lambda_a-\lambda_b)^2 \nonumber \\
& \qquad  \mp (\lambda_a-\lambda_b)\sqrt{(\lambda_a+\lambda_b)^2+\lambda_c^2}\big] \\
K^2=& 8 \big[ (\lambda_a^2-\lambda_b^2) +(\lambda_a-\lambda_b)^2 \lambda_c^2 \nonumber\\
 & +(\lambda_a^2-\lambda_b^2)(\lambda_a-\lambda_b)\sqrt{(\lambda_a+\lambda_b)^2+\lambda_c^2} \big] \\
v_{\Delta \pm} =& 2 (\lambda_a+\lambda_b)^2+\lambda_c^2 \nonumber\\
&\pm  (\lambda_a-\lambda_b)\sqrt{(\lambda_a+\lambda_b)^2+\lambda_c^2}.
\end{align}
\end{subequations} 
Although Eq.~(\ref{eq:uv_special}) has complex prefactors, their square and the cross terms do not depend on ${\bf p}$. Thus, we can take place the integral in terms of ${\bf p}$, ${\bf p}'$ first. One yields
\begin{align}
 \frac{1}{T_1} = & \pi  \gamma_{\rm n}^2 A_{\rm hf}^2  \frac{7 k_F^3}{240v_F^2 v_{\Delta+}^2v_{\Delta-}^2} T^5 \nonumber \\ 
 &\times \Big[ \frac{v_{\Delta -}^2}{v_{\Delta +}^2} D_+^{[3,1,\frac{3}{2}]}(\theta_I,\phi_I) +  \frac{v_{\Delta +}^2}{v_{\Delta -}^2} D_-^{[3,1,\frac{3}{2}]}(\theta_I,\phi_I) \nonumber \\
 &\quad +  D_{+-}^{[3,1,\frac{3}{2}]}(\theta_I,\phi_I) + D_{-+}^{[3,1,\frac{3}{2}]}(\theta_I,\phi_I) \Big]. \label{eq:NMR3132}
\end{align}
Each coefficient depends on $\theta_I$ and $\phi_I$. The anisotropy of coefficient in terms of the nuclear spin direction, i.e., the  magnetic field direction, is visualized in Fig.~\ref{fig:coeff2}. We explicitly show the coefficients as follows:
\begin{widetext}
\begin{subequations} \label{eq:cf_special}
\begin{align}
 &D_+^{[3,1,\frac{3}{2}]}(\theta_I,\phi_I)  \nonumber \\
 &= \left(\frac{1+\cos(\theta_I)^2}{2}- \cos^2 (\theta_I) \sin^2(\theta_I) \right)  \left( \frac{f_x^2(\sqrt{f_x^2 +f_z^2} +f_z)^2}{K^4} + \frac{(g_z^{-})^2(g_x^-+g_y^-)^2}{K^4 v_{\Delta+}^4}\right) \nonumber \\
 &\quad +\frac{\sin^2(\theta_I)}{4} \left( \frac{(\sqrt{f_x^2 +f_z^2} +f_z)^4 +f_x^4}{K^4} +  \frac{(g_x^-+g_y^-)^4+(g_z^{-})^4}{K^4 v_{\Delta+}^4} \right) \nonumber \\
 &\quad -\frac{\cos(\phi_I) \sin(2 \theta_I)}{2} \left( \frac{f_x(\sqrt{f_x^2 +f_z^2} +f_z)^3 -f_x^3(\sqrt{f_x^2 +f_z^2} +f_z)}{K^4} +  \frac{(g_z^{-})^3(g_x^-+g_y^-)-g_z^{-}(g_x^-+g_y^-)^3}{K^4 v_{\Delta+}^4} \right), \\
  &D_-^{[3,1,\frac{3}{2}]}(\theta_I,\phi_I)  \nonumber \\
 &= \left(\frac{1+\cos(\theta_I)^2}{2}- \cos^2 (\theta_I) \sin^2(\theta_I) \right)  \left( \frac{f_x^2(\sqrt{f_x^2 +f_z^2} +f_z)^2}{K^4} + \frac{(g_z^{+})^2(g_x^+-g_y^+)^2}{K^4 v_{\Delta-}^4}\right) \nonumber \\
 &\quad +\frac{\sin^2(\theta_I)}{4} \left( \frac{(\sqrt{f_x^2 +f_z^2} +f_z)^4 +f_x^4}{K^4} +  \frac{(g_x^+-g_y^+)^4+(g_z^{+ })^4}{K^4 v_{\Delta-}^4} \right) \nonumber \\
 &\quad -\frac{\cos(\phi_I) \sin(2 \theta_I)}{2} \left( \frac{f_x(\sqrt{f_x^2 +f_z^2} +f_z)^3 -f_x^3(\sqrt{f_x^2 +f_z^2} +f_z)}{K^4} +  \frac{(g_z^{+})^3(g_x^+-g_y^+)-g_z^{+}(g_x^+-g_y^+)^3}{K^4 v_{\Delta-}^4} \right), \\
  &D_{+-}^{[3,1,\frac{3}{2}]}(\theta_I,\phi_I) + D_{-+}^{[3,1,\frac{3}{2}]}(\theta_I,\phi_I)  \nonumber \\
  &=\frac{1+\cos(\theta_I)^2}{2} \left(\frac{(\sqrt{f_x^2 +f_z^2} +f_z)^4 +f_x^4}{K^4} + \frac{(g_x^-+g_y^-)^2(g_x^+-g_y^+)^2+g_z^{- 2}(g_z^{+})^2}{K^4 v_{\Delta+}^2v_{\Delta+}^2}\right)\nonumber \\
  &\quad +\frac{\sin^2(\theta_I)}{2} \left( \frac{2(\sqrt{f_x^2 +f_z^2} +f_z)^2 f_x^2}{K^4} + \frac{(g_x^-+g_y^-)^2 (g_z^{+})^2+ (g_x^+-g_y^+)^2 (g_z^{-})^2}{K^4 v_{\Delta+}^2v_{\Delta+}^2}\right)  \nonumber \\
  &\quad +2 \cos^2 (\theta_I) \sin^2(\theta_I)  \left( \frac{(\sqrt{f_x^2 +f_z^2} +f_z)^2 f_x^2}{K^4} + \frac{(g_x^-+g_y^-)(g_x^+-g_y^+)g_z^{-}g_z^{+}}{K^4 v_{\Delta+}^2v_{\Delta+}^2}\right)  \nonumber \\
  &\quad -\frac{\cos(\phi_I) \sin(2 \theta_I)}{2}\Bigg( 2\frac{f_x^3(\sqrt{f_x^2 +f_z^2} +f_z) -f_x(\sqrt{f_x^2 +f_z^2} +f_z)^3}{K^4}  \nonumber \\
  &\quad+\frac{g_z^{+}(g_x^-+g_y^-)^2(g_x^+-g_y^+)+g_z^{-}(g_x^-+g_y^-)(g_x^+-g_y^+)^2-g_z^{-}(g_z^{+})^2(g_x^-+g_y^-)-(g_z^{-})^2g_z^{+ }(g_x^+-g_y^+)}{K^4 v_{\Delta+}^2v_{\Delta-}^2}  \Bigg).
\end{align}
\end{subequations} 
\end{widetext}

\bibliography{spin_relaxation}

\end{document}